\documentclass[preprint]{sigplanconf}
\usepackage[varg]{txfonts}

\usepackage{amssymb,amsfonts,amsmath}

\usepackage[mathscr]{eucal}

\usepackage{stmaryrd}

\usepackage{amsbsy}

\usepackage{bm}

\usepackage{braket}

\newcommand{\Tra}{{\sf T}}

\newcommand{\Dsquare}[1]{\left\llbracket #1 \right\rrbracket}

\newcommand{\Real}{\mathbb{R}}

\newcommand{\V}[2][]{{\bm{#1\mathbf{\MakeLowercase{#2}}}}} 
\newcommand{\VE}[3][]{#1{\MakeLowercase{#2}}_{#3}} 
 
\newcommand{\M}[2][]{{\bm{#1\mathbf{\MakeUppercase{#2}}}}} 
 
\newcommand{\MC}[3][]{\V[#1]{#2}_{#3}} 
 
\newcommand{\ME}[3][]{#1{\MakeLowercase{#2}}_{#3}} 

\newcommand{\T}[2][]{\boldsymbol{#1\mathscr{\MakeUppercase{#2}}}} 
\newcommand{\TS}[3][]{\M[#1]{#2}_{#3}}
\newcommand{\TE}[3][]{#1{\MakeLowercase{#2}}_{#3}}

\newcommand{\Oprod}{\circ} 
\newcommand{\Kron}{\otimes} 
 
\newcommand{\Hada}{\ast} 
 
\usepackage{amsmath}
\usepackage[]{algorithm2e, setspace}
\usepackage{amsfonts}
\usepackage{booktabs}
\usepackage{graphicx}
\usepackage{color}
\usepackage[colorlinks]{hyperref}
\usepackage{url}

\newtheorem{theorem}{Theorem}[section]

\newtheorem{proposition}[theorem]{Proposition}

\newcommand{\qed}{\nobreak \ifvmode \relax \else
      \ifdim\lastskip<1.5em \hskip-\lastskip
      \hskip1.5em plus0em minus0.5em \fi \nobreak
      \vrule height0.75em width0.5em depth0.25em\fi}

\begin{document}

\newcommand{\TODO}[1]{\textcolor{blue}{\textbf{TODO XXXXX: #1}}}
\newcommand{\GB}[1]{\textcolor{red}{\textbf{GB: #1}}}
\newcommand{\vecmat}[1]{\text{vec}\left(#1\right)}
\newcommand{\nnz}[1]{\text{nnz}\left(#1\right)}
\newcommand{\diag}[1]{\text{diag}\left(#1\right)}
\newcommand{\bc}[3]{\left\langle#1, #2, #3\right\rangle}
\newcommand{\alg}[3]{\Dsquare{#1, #2, #3}}
\newcommand{\dims}[3]{#1 \times #2 \times #3}
\newcommand{\daxpy}{\texttt{daxpy}}
\newcommand{\dgemm}{\texttt{dgemm}}
\newcommand{\lt}{\left}
\newcommand{\rt}{\right}

\setlength{\pdfpageheight}{\paperheight}
\setlength{\pdfpagewidth}{\paperwidth}

\conferenceinfo{PPoPP '15}{Month d--d, 2015, San Francisco, CA, USA}
\copyrightyear{2015} 
\copyrightdata{978-1-nnnn-nnnn-n/yy/mm} 
\doi{nnnnnnn.nnnnnnn}




\titlebanner{}
\preprintfooter{}

\title{A Framework for Practical Parallel Fast Matrix Multiplication}


\authorinfo{Austin R. Benson}
           {Stanford University}
           {arbenson@stanford.edu}
\authorinfo{Grey Ballard}
           {Sandia National Laboratories}
           {gmballa@sandia.gov}

\maketitle

\begin{abstract}
Matrix multiplication is a fundamental computation in many scientific disciplines.
In this paper, we show that novel fast matrix multiplication algorithms can significantly outperform vendor implementations of the classical algorithm and Strassen's fast algorithm on modest problem sizes and shapes.
Furthermore, we show that the best choice of fast algorithm depends not only on the size of the matrices but also the shape.
We develop a code generation tool to automatically implement multiple sequential and shared-memory parallel variants of each fast algorithm, including our novel parallelization scheme.
This allows us to rapidly benchmark over 20 fast algorithms on several problem sizes.
Furthermore, we discuss a number of practical implementation issues for these algorithms on shared-memory machines that can direct further research on making fast algorithms practical.
\end{abstract}

\category{G.4}{Mathematical software}{Efficiency}
\category{G.4}{Mathematical software}{Parallel and vector implementations}


\keywords
fast matrix multiplication, dense linear algebra, parallel linear algebra, shared memory

\section{Introduction}

Matrix multiplication is one of the most fundamental computations in numerical linear algebra and scientific computing.
Consequently, the computation has been extensively studied in parallel computing environments \cite{van1997summa, solomonik2011communication, irony2004communication, ballard2012communication, volkov2008benchmarking}.
In this paper, we show that fast algorithms for matrix-matrix multiplication can achieve higher performance on sequential and shared-memory parallel architectures for modestly sized problems.
By fast algorithms, we mean ones that perform asymptotically fewer floating point operations and communicate asymptotically less data than the classical algorithm.
We also provide a code generation framework to rapidly implement sequential and parallel versions of over 20 fast algorithms.
Our performance results in Section~\ref{sec:performance} show that several fast algorithms can outperform the Intel Math Kernel Library (MKL) $\dgemm$ (double precision general matrix-matrix multiplication) routine and Strassen's algorithm~\cite{strassen1969gaussian}.
In parallel implementations, fast algorithms can achieve a speedup of 5\% over Strassen's algorithm and greater than 15\% over MKL.

However, fast algorithms for matrix multiplication have largely been ignored in practice.
For example, numerical libraries such as Intel's MKL, AMD's Core Math Library (ACML),
and the Cray Scientific Libraries package (LibSci) do not provide implementations of fast algorithms.
Why is this the case?
First, users of numerical libraries typically consider fast algorithms to be of only theoretical interest and never practical for reasonable problem sizes.
We argue that this is \emph{not} the case with our performance results in Section~\ref{sec:performance}.
Second, fast algorithms do not provide the same numerical stability guarantees as the classical algorithm.
In practice, there is some loss in precision in the fast algorithms, but they are not nearly as bad as the worst-case guarantees \cite{higham2002accuracy,lipshitz2012communication}.
Third, the LINPACK benchmark used to rank supercomputers by performance forbids fast algorithms \cite{TOP500}.
We suspect that this has driven effort away from the study of fast algorithms.

Strassen's algorithm is the most well known fast algorithm, but this paper explores a much larger class of recursive fast algorithms based on different base case dimensions.
We review these algorithms and methods for constructing them in Section~\ref{sec:fast}.
The structure of these algorithms makes them amenable to code generation, and we describe this process and other performance tuning considerations in Section~\ref{sec:implementation}.
In Section~\ref{sec:parallel}, we describe three different methods for parallelizing fast matrix multiplication algorithms on shared-memory machines.
Our code generator implements all three parallel methods for each fast algorithm.
We evaluate the sequential and parallel performance characteristics of the various algorithms and implementations in Section~\ref{sec:performance} and compare them with MKL's implementation of the classical algorithm as well as an existing implementation of Strassen's algorithm.

The goal of this paper is to help bridge the gap between theory and practice of fast matrix multiplication algorithms.
By introducing our tool of automatically translating a fast matrix multiplication algorithm to high performance sequential and parallel implementations, we enable the rapid prototyping and testing of theoretical developments in the search for faster algorithms.
We focus the attention of theoretical researchers on what algorithmic characteristics matter most in practice, and we demonstrate to practical researchers the utility of several existing fast algorithms besides Strassen's, motivating further effort towards high performance implementations of those that are most promising.

Our contributions are summarized as follows:
\begin{itemize}
\item 
By using new fast matrix multiplication algorithms, we achieve better performance than Intel MKL's $\dgemm$, both sequentially and with 6 and 24 cores on a shared-memory machine.

\item 
We demonstrate that, in order to achieve the best performance for matrix multiplication, the choice of fast algorithm depends on the size and shape of the matrices.
Our new fast algorithms outperform Strassen's algorithm on the multiplication of rectangular matrices.

\item
We show how to use code generation techniques to rapidly implement fast matrix multiplication algorithms.

\item
We provide a new hybrid parallel algorithm for shared-memory fast matrix multiplication.

\item
We implement a fast matrix multiplication algorithm with asymptotic complexity $O(N^{2.775})$ for square $N \times N$ matrices.
In terms of asymptotic complexity, this is the fastest matrix multiplication algorithm implementation to date.
However, our performance results show that this algorithm is not practical for the problem sizes that we consider.


\end{itemize}

Overall, we find that Strassen's algorithm is hard to beat for square matrix multiplication, both in serial and in parallel.
However, for rectangular matrices (which occur more frequently in practice), other fast algorithms can perform much better.
The structure of the fast algorithms that perform well tend to ``match the shape" of the matrices, an idea that we will make clear in Section~\ref{sec:performance}.
We also find that bandwidth is a factor towards scalability in shared-memory parallel implementations of fast algorithms.
Finally, we find that algorithms that are theoretically fast in terms of asymptotic complexity do not perform well on problems of modest size that we consider on shared-memory parallel architectures.

All of the software used for this paper will be made publicly available.

\subsection{Related work}
\label{sec:related}

Strassen's fast matrix multiplication algorithm has been implemented for both shared-memory~\cite{kumar1995tensor, d2011exploiting} and distributed-memory architectures~\cite{grayson1996high, luo1995scalable, ballard2012communication}.
For our parallel algorithms in Section~\ref{sec:parallel}, we use the ideas of breadth-first and depth-first traversals of the recursion trees, which were first considered by Kumar \emph{et al}.~\cite{kumar1995tensor} and Ballard \emph{et al}.~\cite{ballard2012communication} for minimizing memory footprint and communication.

Apart from Strassen's algorithm, a number of fast matrix multiplication algorithms have been developed,
but only a small handful have been implemented.
Furthermore, these implementations have only been sequential.
Hopcroft and Kerr showed how to construct recursive fast algorithms where the base case is multiplying a $p \times 2$ by a $2 \times n$ matrix~\cite{hopcroft1971minimizing}.
Bini \emph{et al}. introduced the concept of arbitrary precision approximate (APA) algorithms for matrix multiplication and demonstrated a method for multiplying $3\times2$ by $2\times2$ matrices which leads to a general square matrix multiplication APA algorithm that is asymptotically faster than Strassen's \cite{BCRL79}.
Sch\"{o}nhage also developed an APA algorithm that is asymptotically faster than Strassen's, based on multiplying $3\times3$ by $3\times3$ matrices \cite{Schonhage81}.
These APA algorithms suffer from severe numerical issues---both lose at least half the digits of accuracy with each recursive step.
While no exact solution can have the same complexity as Bini's algorithm \cite{hopcroft1971minimizing},
it is still an open question if there exists a fast algorithm with the same complexity as Sch\"{o}nhage's.
Pan used factorization of trilinear forms and a base case of $70 \times 70$ matrix multiplication to construct an exact algorithm asymptotically faster than Strassen's algorithm~\cite{pan1978strassen}.
This algorithm was implemented by Kaporin~\cite{kaporin2004aggregation}, and the running time was competitive with Strassen's algorithm in practice.
Recently, Smirnov presented optimization tools for finding many fast algorithms based on factoring bilinear forms~\cite{smirnov2013bilinear},
and we will use these tools for finding our own algorithms in Section~\ref{sec:fast}.
Other automated approaches have also been used to discover fast algorithms, but these have focused on multiplying $3\times 3$ by $3\times 3$ matrices \cite{JM86,CBH11,OKM13}.

There are several lines of theoretical research~\cite{coppersmith1990matrix, stothers2010complexity, williams2012multiplying} that prove existence of fast APA algorithms with much better asymptotic complexity than the algorithms considered in this paper.
Unfortunately, there is still a large gap between the substantial theoretical work and what we can practically implement.

Renewed interest in the practicality of Strassen's and other fast algorithms is motivated by the observation that not only is the arithmetic cost reduced when compared to the classical algorithm, the communication costs also improve asymptotically \cite{ballard2012graph}.
That is, as the relative cost of moving data throughout the memory hierarchy and between processors increases, we can expect the benefits of fast algorithms to grow accordingly.
We note that communication lower bounds (\cite[Theorem 1.4]{ballard2012graph} and \cite[Theorem 1]{BDHLS12-RECT}) apply to all the algorithms presented in this paper, and in nearly all cases they are attained by the implementations used in this paper.

\subsection{Notation and tensor preliminaries}

The relevant notation for our work is in Table~\ref{tab:notation}.
Throughout, scalars are represented by lowercase Roman or Greek letters ($a$), vectors by lowercase boldface ($\V{x}$),
matrices by uppercase boldface ($\M{A}$), and tensors by boldface Euler script letters ($\T{T}$).
For a matrix $\M{A}$, we use $\MC{A}{k}$ and $\ME{A}{ij}$ to denote the $k$th column and $i, j$ entry, respectively.
We briefly review basic tensor preliminaries, following the notation of Kolda and Bader \cite{kolda2009tensor}.
A tensor is a multi-dimensional array, and in this paper we deal exclusively with order-3, real-valued tensors; \emph{i.e.}, $\T{T} \in \Real^{I \times J \times K}$.
The $k$th \emph{frontal slice} of $\T{T}$ is $\TS{T}{k} = \TE{T}{:,:,k} \in \Real^{I \times J}$.
For $\V{u} \in \Real^I$, $\V{v} \in \Real^J$, $\V{w} \in \Real^K$, we define the outer product tensor
$\T{T} = \V{u} \Oprod \V{v} \Oprod \V{w} \in  \Real^{I \times J \times K}$ with entries $\TE{T}{ijk} = \VE{u}{i}\VE{v}{j}\VE{w}{k}$.
Addition of tensors is defined entry-wise.
The \emph{rank} of a tensor $\T{T}$ is the minimum number of rank-one tensors that generate $\T{T}$ as their sum.
Decompositions of the form $\T{T} = \sum_{r=1}^{R}\V{u}_r \Oprod \V{v}_r \Oprod \V{w}_r$ lead to fast matrix multiplication algorithms (Section~\ref{sec:tensor_fact}), and we use $\alg{\M{U}}{\M{V}}{\M{W}}$ to denote the decomposition, where $\M{U}$, $\M{V}$, and $\M{W}$ are matrices with $R$ columns given by $\V{u}_r$, $\V{v}_r$, and $\V{w}_r$.
Of the various flavors of products involving tensors, we will need to know that, for $\V{a} \in \Real^{I}$ and $\V{b} \in \Real^{J}$,
$\T{T} \times_{1} \V{a} \times_{2} \V{b} = \V{c} \in \Real^{K}$, with $\VE{c}{k}=\V{a}^{\Tra}\TS{T}{k}\V{b}$, or $\VE{c}{k}=\sum_{i=1}^I\sum_{j=1}^J \TE{T}{ijk}\VE{a}{i}\VE{b}{j}$.

\begin{table}[tb]
\centering
\caption{
Summary of notation.
}
\begin{tabular}{l l}
\toprule \\
$\bc{M}{K}{N}$ & ``base case'' computation, multiplying an $M \times K$  \\
& matrix by a $K \times N$ matrix \\
$\dims{P}{Q}{R}$ & dimensions of actual matrices that are multiplied \\
& ($P \times Q$ matrix multiplied by $Q \times R$ matrix) \\
$\alg{\M{U}}{\M{V}}{\M{W}}$ & factor matrices corresponding to a tensor \\
& decomposition that provides a fast algorithm \\
$\T{T}$ & order three tensor \\
$\T{T} = u \Oprod v \Oprod w$ & rank-1 tensor with entries $\ME{T}{ijk} = u_iv_jw_k$ \\
$\M{T}_i$ & $i$th frontal slice of $\T{T}$, the matrix of entries $\TE{T}{:,:,i}$ \\
$\MC{A}{k}$, $\ME{A}{ij}$ & $k$th column and $i, j$ entry of matrix $\M{A}$ \\
$\vecmat{\M{A}}$ & row-order vectorization of the entries of $\M{A}$ \\
$\nnz{\cdot}$ & number of non-zero entries of an object \\
\bottomrule
\end{tabular}
\label{tab:notation}
\end{table}

\section{Fast matrix multiplication}
\label{sec:fast}

We now review the preliminaries for fast matrix multiplication algorithms.
In particular, we focus on factoring tensor representations of bilinear forms,
which will facilitate the discussion of the implementation in Sections~\ref{sec:implementation}~and~\ref{sec:parallel}.

\subsection{Recursive multiplication}
\label{sec:recursive}

Matrices are self-similar, \emph{i.e.}, a submatrix is also a matrix.
Arithmetic with matrices is closely related to arithmetic with scalars, and
we can build recursive matrix multiplication algorithms by manipulating submatrix blocks.
For example, consider multiplying $C = A \cdot B$,
\[
\begin{bmatrix}
\M{C}_{11} & \M{C}_{12} \\
\M{C}_{21} & \M{C}_{22}
\end{bmatrix} = 
\begin{bmatrix}
\M{A}_{11} & \M{A}_{12} \\
\M{A}_{21} & \M{A}_{22}
\end{bmatrix} \cdot
\begin{bmatrix}
\M{B}_{11} & \M{B}_{12} \\
\M{B}_{21} & \M{B}_{22}
\end{bmatrix},
\]
where we have partitioned the matrices into four submatrices.
Throughout this paper, we denote the block multiplication of $M \times K$ and $K \times N$ matrices by $\bc{M}{K}{N}$.
Thus, the above computation is $\bc{2}{2}{2}$.
Multiplication with the classical algorithm proceeds by combining a set of eight matrix multiplications with four matrix additions:
\begin{align*}
\M{M}_1 &= \M{A}_{11} \cdot \M{B}_{11}  & \M{M}_2 &= \M{A}_{12} \cdot \M{B}_{21} & \M{M}_3 &= \M{A}_{11} \cdot \M{B}_{12} \\
\M{M}_4 &= \M{A}_{12} \cdot \M{B}_{22} & \M{M}_5 &= \M{A}_{21} \cdot \M{B}_{11} & \M{M}_6 &= \M{A}_{22} \cdot \M{B}_{21} \\
\M{M}_7 &= \M{A}_{21} \cdot \M{B}_{12} & \M{M}_8 &= \M{A}_{22} \cdot \M{B}_{22}
\end{align*}
\vspace{-0.7cm} 
\begin{align*}
\M{C}_{11} &= \M{M}_1 + \M{M}_2 & \M{C}_{12} &= \M{M}_3 + \M{M}_4 \\
\M{C}_{21} &= \M{M}_5 + \M{M}_6  & \M{C}_{22} &= \M{M}_7 + \M{M}_8
\end{align*}
The multiplication to form each $\M{M}_i$ is recursive and the base case is scalar multiplication.
The number of flops performed by the classical algorithm for $N \times N$ matrices, where $N$ is a power of two, is:
\[
   F_C(N) = \left\{
     \begin{array}{lr}
       8F_C(N / 2) + 4(N / 2)^2 & : N > 1 \\
       1 & : N = 1
     \end{array}
   \right.
\]
This is a standard recurrence relation with $F_C(N) = 2N^3 -N^2$.
We have assumed that the matrices are square and powers of two.
In Section~\ref{sec:all_dimensions}, we explain how to handle all dimensions.

The idea of fast matrix multiplication algorithms is to perform fewer recursive matrix multiplications at the expense of more matrix additions.
Since matrix multiplication is asymptotically more expensive than matrix addition, this tradeoff results in faster algorithms.
The most well known fast algorithm is due to Strassen, and follows the same block structure:
\begin{align*}
\M{S}_1 &= \M{A}_{11} + \M{A}_{22} & \M{S}_2 &= \M{A}_{21} + \M{A}_{22} & \M{S}_3 &= \M{A}_{11} \\
\M{S}_4 &= \M{A}_{22} & \M{S}_5 &= \M{A}_{11} + \M{A}_{12} & \M{S}_6 &= \M{A}_{21} - \M{A}_{11} \\
\M{S}_7 &= \M{A}_{12} - \M{A}_{22} \\
\M{T}_1 &= \M{B}_{11} + \M{B}_{22} & \M{T}_2 &= \M{B}_{11} & \M{T}_3 &= \M{B}_{12} - \M{B}_{22} \\
\M{T}_4 &= \M{B}_{21} - \M{B}_{11} & \M{T}_5 &= \M{B}_{22} &\M{T}_6 &= \M{B}_{11} + \M{B}_{12} \\
\M{T}_7 &= \M{B}_{21} + \M{B}_{22}
\end{align*}
\begin{align*}
\M{M}_r &= \M{S}_r\M{T}_r, \quad 1 \le r \le 7 \\
\M{C}_{11} &= \M{M}_1 + \M{M}_4 - \M{M}_5 + \M{M}_7 & \M{C}_{12} &= \M{M}_3 + \M{M}_5 \\
\M{C}_{21} &= \M{M}_2 + \M{M}_4 & \M{C}_{22} &= \M{M}_1 - \M{M}_2 + \M{M}_3 + \M{M}_6
\end{align*}

We have explicitly written out terms like $\M{T}_2 = \M{B}_{11}$ to hint at the generalizations provided in Section~\ref{sec:tensor_fact}.
Strassen's algorithm uses 7 matrix multiplications and 18 matrix additions.
The number of flops performed by the algorithm is:
\[
   F_S(N) = \left\{
     \begin{array}{lr}
       7F_C(N / 2) + 18(N / 2)^2 & : N > 1 \\
       1 & : N = 1
     \end{array}
   \right.
\]
and $F_S(N) = 7N^{\log_2 7} - 6N^2 = O(N^{2.81})$.

There are natural extensions to Strassen's algorithm.
We might try to find an algorithm using fewer than 7 multiplications; unfortunately, we cannot~\cite{winograd1971multiplication}.
Alternatively, we could try to reduce the number of additions.
This leads to the Strassen-Winograd algorithm, which reduces the 18 additions down to 15, which is also optimal \cite{Probert76}.
We explore such methods in Section~\ref{sec:cse}.
We can also improve the constant on the leading term by choosing a bigger base case dimension (and using the classical algorithm for the base case).
This turns out not to be important in practice because the base case will be chosen to optimize performance rather than flop count.
Lastly, we can use blocking schemes apart from $\bc{2}{2}{2}$, which we explain in the remainder of this section.
This leads to a host of new algorithms, and we show in Section~\ref{sec:performance} that they are often faster in practice.

\subsection{Fast algorithms as low-rank tensor decompositions}
\label{sec:tensor_fact}

The approach we use to devise fast algorithms exploits an important connection between matrix multiplication (and other bilinear forms) and tensor computations.
We detail the connection in this section for completeness; see \cite{Brent70,Kruskal77,Knuth81} for earlier explanations.


A bilinear form on a pair of finite-dimensional vector spaces is a function that maps a pair of vectors to a scalar and is linear in each of its inputs separately.
A bilinear form $B(\V{x},\V{y})$ can be represented by a matrix $\M{M}$ of coefficients: $B(\V{x},\V{y}) = \V{x}^{\Tra}\M{M}\V{y} = \sum_i \sum_j \ME{M}{ij} \VE{x}{i} \VE{y}{j}$, where we note that $\V{x}$ and $\V{y}$ may have different dimensions.
In order to describe a set of $K$ bilinear forms $B_k(\V{x},\V{y})=\VE{z}{k}$, $1 \le k \le K$, we can use a three-way tensor $\T{T}$ of coefficients:
\begin{equation}
\label{eqn:conventional}
\VE{z}{k} = \sum_{i=1}^I\sum_{j=1}^J \TE{T}{ijk} \VE{x}{i} \VE{y}{j},
\end{equation}
or, in more succinct tensor notation, $\V{z} = \T{T} \times_{1} \V{x} \times_{2} \V{y}$.

\subsubsection{Low-rank tensor decompositions}

The advantage of representing the operations using a tensor of coefficients is a key connection between the rank of the tensor to the arithmetic complexity of the corresponding operation.
Consider the ``active'' multiplications between elements of the input vectors (\emph{e.g.}, $\VE{x}{i} \cdot \VE{y}{j}$).
The conventional algorithm, following Equation \eqref{eqn:conventional}, will compute an active multiplication for every nonzero coefficient in $\T{T}$.
However, suppose we have a rank-$R$ decomposition of the tensor, $\T{T} = \sum_{i=1}^{R} u_i \Oprod v_i \Oprod w_i$, so that
\begin{equation}
\label{eqn:lowrank}
\TE{T}{ijk} = \sum_{r=1}^R \ME{u}{ir} \ME{v}{jr} \ME{w}{kr}
\end{equation}
for all $i$, $j$, $k$, where $\M{U}$, $\M{V}$, and $\M{W}$ are matrices with $R$ columns each.
We will also use the equivalent notation $\T{T}=\alg{\M{U}}{\M{V}}{\M{W}}$.
Substituting Equation \eqref{eqn:lowrank} into Equation \eqref{eqn:conventional} and rearranging, we have for $k=1,\dots,K$,
\begin{align*}
\VE{z}{k} &= \sum_{i=1}^I\sum_{j=1}^J \lt(\sum_{r=1}^R \ME{u}{ir} \ME{v}{jr} \ME{w}{kr}\rt) \VE{x}{i} \cdot \VE{y}{j} \\
& = \sum_{r=1}^R \lt(\sum_{i=1}^I \ME{u}{ir}\VE{x}{i}\rt) \cdot \lt(\sum_{j=1}^J \ME{v}{jr}\VE{y}{j}\rt) \ME{w}{kr} \\
& = \sum_{r=1}^R \lt(\VE{s}{r} \cdot \VE{t}{r} \rt) \ME{w}{kr} = \sum_{r=1}^R \VE{m}{r} \ME{w}{kr},
\end{align*}
which reduces the number of active multiplications (now between linear combinations of elements of the input vectors) to $R$.
Here we highlight active multiplications with $(\cdot)$ notation, $\V{s}$ and $\V{t}$ are temporary vectors that store the linear combinations of elements of $\V{x}$ and $\V{y}$, and $\V{m}$ is the temporary vector that stores the element-wise product of $\V{s}$ and $\V{t}$.  
In matrix-vector notation, we have $\V{s}=\M{U}^{\Tra}  \V{x}$, $\V{t}=\M{V}^{\Tra}  \V{y}$, and $\V{m}=\V{s} \Hada \V{t}$, where ($\Hada$) denotes element-wise vector multiplication.

Assuming $R<\nnz{\T{T}}$, this reduction of active multiplications, at the expense of increasing the number of other operations, is valuable when active multiplications are much more expensive than the other operations.
This is the case for recursive matrix multiplication, where the elements of the input vectors are (sub)matrices, as we describe below.

\subsubsection{Tensor representation of matrix multiplication}

Matrix multiplication is a bilinear operation, so we can represent it as a tensor computation. 
In order to match the notation above, we vectorize the input and output matrices $\M{A}$, $\M{B}$, and $\M{C}$ using row-wise ordering, so that $\V{x}=\vecmat{\M{A}}$, $\V{y}=\vecmat{\M{B}}$, and $\V{z}=\vecmat{\M{C}}$. 

For every triplet of matrix dimensions for valid matrix multiplication, there is a fixed tensor that represents the computation so that $\T{T} \times_{1} \vecmat{\M{A}} \times_{2} \vecmat{\M{B}} = \vecmat{\M{C}}$ holds for all $\M{A}$, $\M{B}$, and $\M{C}$.
For example, if $\M{A}$ and $\M{B}$ are both $2\times 2$, the corresponding $4\times4\times4$ tensor $\T{T}$ has frontal slices
\begin{align*}
\M{T}_1 &= \begin{bmatrix}
1 & 0 & 0 & 0 \\
0 & 0 & 1 & 0 \\
0 & 0 & 0 & 0 \\
0 & 0 & 0 & 0 \\
\end{bmatrix} &
\M{T}_2 = \begin{bmatrix}
0 & 1 & 0 & 0 \\
0 & 0 & 0 & 1 \\
0 & 0 & 0 & 0 \\
0 & 0 & 0 & 0 \\ 
\end{bmatrix} \phantom{.} \\
\M{T}_3 &= \begin{bmatrix}
0 & 0 & 0 & 0 \\
0 & 0 & 0 & 0 \\
1 & 0 & 0 & 0 \\
0 & 0 & 1 & 0 \\
\end{bmatrix} &
\M{T}_4 = \begin{bmatrix}
0 & 0 & 0 & 0 \\
0 & 0 & 0 & 0 \\
0 & 1 & 0 & 0 \\
0 & 0 & 0 & 1 \\
\end{bmatrix}.
\end{align*}
This yields, for example,
\begin{align*}
& \M{T}_3 \times_{1} \vecmat{\M{A}} \times_{2} \vecmat{\M{B}} \\
&= \begin{bmatrix} \ME{A}{11} & \ME{A}{12} & \ME{A}{21} & \ME{A}{22} \end{bmatrix}
\begin{bmatrix}
0 & 0 & 0 & 0 \\
0 & 0 & 0 & 0 \\
1 & 0 & 0 & 0 \\
0 & 0 & 1 & 0 \\
\end{bmatrix}
\begin{bmatrix} \ME{B}{11} \\ \ME{B}{12} \\ \ME{B}{21} \\ \ME{B}{22} \end{bmatrix} \\
&= \ME{A}{21}\cdot \ME{B}{11} + \ME{A}{22} \cdot \ME{B}{21}
= \ME{C}{21}
\end{align*}

By Strassen's algorithm, we know that although this tensor has 8 nonzero entries, its rank is at most 7.
Indeed, that algorithm corresponds to a low-rank decomposition represented by the following triplet of matrices, each with 7 columns:
\begin{align*}
\M{U} &= \begin{bmatrix}
1 & 0 & 1 & 0 & 1 & -1 & 0 \\
0 & 0 & 0 & 0 & 1 & 0 & 1 \\
0 & 1 & 0 & 0 & 0 & 1 & 0 \\
1 & 1 & 0 & 1 & 0 & 0 & -1
\end{bmatrix} \\
\M{V} &= \begin{bmatrix}
1 & 1 & 0 & -1 & 0 & 1 & 0 \\
0 & 0 & 1 & 0 & 0 & 1 & 0 \\
0 & 0 & 0 & 1 & 0 & 0 & 1 \\
1 & 0 & -1 & 0 & 1 & 0 & 1 \\
\end{bmatrix} \\
\M{W} &= \begin{bmatrix}
1 & 0 & 0 & 1 & -1 & 0 & 1 \\
0 & 1 & 0 & 1 & 0 & 0 & 0 \\
0 & 0 & 1 & 0 & 1 & 0 & 0 \\
1 & -1 & 1 & 1 & 0 & 0 & 0 \\
\end{bmatrix}.
\end{align*}
And, as in the previous section, for example,
\begin{align*}
\VE{S}{1} &= \MC{u}{1}^{\Tra}\vecmat{\M{A}} = \ME{A}{11} + \ME{A}{22} \\
\VE{T}{1} &=  \MC{v}{1}^{\Tra}\vecmat{\M{B}} = \ME{B}{11} + \ME{B}{22} \\
\ME{C}{11} &=  \MC{e}{1}^{\Tra}\M{W} \V{m} = \VE{M}{1} + \VE{M}{4} - \VE{M}{5} + \VE{M}{7}.
\end{align*}
Note that in the previous section, the elements of the input matrices are already interpreted as submatrices (\emph{e.g.}, $\M{A}_{11}$ and $\M{M}_1$); here we represent them as scalars (\emph{e.g.}, $a_{11}$ and $m_1$).


We need not restrict ourselves to the $\bc{2}{2}{2}$ case; there exists a tensor for matrix multiplication given any set of valid dimensions.
When considering a base case of $M \times K$ by $K \times N$ matrix multiplication (denoted $\bc{M}{K}{N}$),
the tensor has dimensions $MK \times KN \times MN$ and $MKN$ non-zeros.
In particular, $\TE{T}{ijk}=1$ if the following three conditions hold 
\begin{align*}
(i-1) \mod K &= \lfloor(j-1) / N \rfloor \\
(j-1) \mod N &= (k-1) \mod N \\
\lfloor (i-1) / K \rfloor &= \lfloor (k-1) / N \rfloor 
\end{align*}
and otherwise $\TE{T}{ijk}=0$ (here we assume entries are 1-indexed).

\subsubsection{Approximate tensor decompositions}
The APA algorithms discussed in Section~\ref{sec:related} arise from \emph{approximate} tensor decompositions.
With Bini's algorithm, for example, the factor matrices have entries $1 / \lambda$ and $\lambda$.
As $\lambda \to 0$, the low-rank tensor approximation approaches the true tensor.
However, as $\lambda$ gets small, we suffer from loss of precision in the floating point calculations of the resulting fast algorithm.
Setting $\lambda = \sqrt{\epsilon}$ minimizes the loss of accuracy in Bini's algorithm, where $\epsilon$ is machine precision.




\subsection{Finding fast algorithms}
\label{sec:finding}

We conclude this section with a description of a method for searching for and discovering fast algorithms for matrix multiplication.
Our search goal is to find low-rank decompositions of tensors corresponding to matrix multiplication of a particular set of dimensions, which will identify fast, recursive algorithms with reduced arithmetic complexity. 
That is, given a particular base case $\bc{M}{K}{N}$ and the associated tensor $\T{T}$, we seek a rank $R$ and matrices $\M{U}$, $\M{V}$, and $\M{W}$ that satisfy Equation \eqref{eqn:conventional}.
Table~\ref{tab:algorithms} summarizes the algorithms that we find and use for numerical experiments in Section~\ref{sec:performance}.

The rank of the decomposition determines the number of active multiplications, or recursive calls, and therefore the exponent in the arithmetic cost of the algorithm.
The number of other operations (additions and inactive multiplications) will affect only the constants in the arithmetic cost.
For this reason, we would like to have sparse $\M{U}$, $\M{V}$, and $\M{W}$ matrices with simple values (like $\pm 1$), but that goal is of secondary importance compared to minimizing the rank $R$.
Note that these constant values do affect performance of these algorithms for reasonable matrix dimensions in practice, though mainly because of how they affect the communication costs of the implementations rather than the arithmetic cost.
We discuss this in more detail in Section~\ref{sec:matrix_additions}.

\subsubsection{Equivalent algorithms}

Given an algorithm $\alg{\M{U}}{\M{V}}{\M{W}}$ for base case $\bc{M}{K}{N}$, we can transform it to an algorithm for any of the other 5 permutations of the base case dimensions with the same number of multiplications.
This is a well known property \cite{HM73}; here we state the two transformations that generate all permutations in our notation.
We let $\M{P}_{I\times J}$ be the permutation matrix that swaps row-order for column-order in the vectorization of an $I \times J$ matrix.
In other words, if $\M{A}$ is $I \times J$, $\M{P}_{I\times J} \cdot \vecmat{\M{A}} = \vecmat{\M{A}^{\Tra}}$.

\begin{proposition}
\label{prop:perm1}
Given a fast algorithm $\alg{\M{U}}{\M{V}}{\M{W}}$ for $\bc{M}{K}{N}$, $\alg{\M{P}_{K\times N}\M{V}}{\M{P}_{M\times K}\M{U}}{\M{P}_{M\times N}\M{W}}$ is a fast algorithm for $\bc{N}{K}{M}$. 
\end{proposition}

\begin{proposition}
\label{prop:perm2}
Given a fast algorithm $\alg{\M{U}}{\M{V}}{\M{W}}$ for $\bc{M}{K}{N}$,
$\alg{\M{P}_{M\times N}\M{W}}{\M{U}}{\M{P}_{K\times N}\M{V}}$ is a fast algorithm for $\bc{N}{M}{K}$.
\end{proposition}


We also point out that fast algorithms for a given base case belong to equivalence classes.
Two algorithm are equivalent if one can be generated from another based on the following transformations \cite{deGroote78a,JM86}.

\begin{proposition}
\label{prop:eqalgs}
If $\alg{\M{U}}{\M{V}}{\M{W}}$ is a fast algorithm for $\bc{M}{K}{N}$, then the following are also fast algorithms for $\bc{M}{K}{N}$:
\[
\alg{\M{U}\M{P}}{\M{V}\M{P}}{\M{W}\M{P}}
\]
for any permutation matrix $\M{P}$;
\[
\alg{\M{U}\M{D}_x}{\M{V}\M{D}_y}{\M{W}\M{D}_z}
\]
for any diagonal matrices $\M{D}_x$, $\M{D}_y$, and $\M{D}_z$ such that $\M{D}_x\M{D}_y\M{D}_z=\M{I}$;
\[
\alg{(\M{Y}^{-\Tra}\Kron\M{X})\M{U}}{(\M{Z}^{-\Tra}\Kron\M{Y})\M{V}}{(\M{R}\Kron\M{P}^{-\Tra})\M{W}}
\]
for any nonsingular matrices $\M{X}\in\Real^{M\times M}$, $\M{Y}\in\Real^{K\times K}$, $\M{Z}\in\Real^{N\times N}$.
\end{proposition}

\subsubsection{Numerical search}

Given a rank $R$ for base case $\bc{M}{K}{N}$, Equation \eqref{eqn:conventional} defines $(MKN)^2$ polynomial equations of the form given in Equation \eqref{eqn:lowrank}.
Because the polynomials are trilinear, alternating least squares (ALS) can be used to iteratively compute an approximate (numerical) solution to the equations.
That is, if two of the three factor matrices are fixed, the optimal 3rd factor matrix is the solution to a linear least squares problem.
Thus, each outer iteration of ALS involves alternating among solving for $\M{U}$, $\M{V}$, and $\M{W}$, each of which can be done efficiently with the QR decomposition.
This approach was first proposed for fast matrix multiplication search by Brent \cite{Brent70}, but ALS has been a popular method for general low-rank tensor approximation for as many years (see \cite{kolda2009tensor} and references therein).

The main difficulties ALS faces for this problem include getting stuck at local minima, encountering ill-conditioned linear least-squares problems, and, even if ALS converges to machine-precision accuracy, computing dense $\M{U}$, $\M{V}$, and $\M{W}$ matrices with floating point entries.
We follow the work of Johnson and McLoughlin \cite{JM86} and Smirnov \cite{smirnov2013bilinear} in addressing these problems.
We use multiple starting points to handle the problem of local minima, add regularization to help with the ill-conditioning, and encourage sparsity in order to recover exact factorizations (with integral or rational values) from the approximations.

The most useful techniques in our search have been (1) exploiting the transformations given in Proposition~\ref{prop:eqalgs} to encourage sparsity and obtain discrete values and (2) using and adjusting the regularization penalty term \cite[Equations (4-5)]{smirnov2013bilinear} throughout the iteration. 
As described in earlier efforts, algorithms for small base cases can be discovered nearly automatically.
However, as the values $M$, $N$, and $K$ grow, more hands-on tinkering using heuristics seems to be necessary to find discrete solutions.

\begin{table}[tb]
\centering
\caption{
Summary of fast algorithms.
Algorithms without citation were found by the authors using the ideas in Section~\ref{sec:finding}.
An asterisk denotes an approximation (APA) algorithm.
The number of multiplications  is equal to the rank $R$ of the corresponding tensor decomposition.
The multiplication speedup per recursive step is the expected speedup if
matrix additions were free.
Note that this speedup does not determine the fastest algorithm because the maximum number
of recursive steps depend on the size of the subproblems created by the algorithm.
By Propositions~\ref{prop:perm1}~and~\ref{prop:perm2}, we also have fast algorithms for all permutations of the base case $\bc{M}{K}{N}$.
}
\begin{tabular}{l c c c c}
\toprule
                & Number of  & Number of    & Multiplication \\
Algorithm              & multiplies    &  multiplies     &  speedup per \\
base case &   (fast)         & (classical)     & recursive step \\\midrule
$\bc{2}{2}{3}$ & 11 & 12 & 9\% \\
$\bc{2}{2}{5}$ & 18 & 20 & 11\% \\
$\bc{2}{2}{2}$ \cite{strassen1969gaussian} & 7 & 8 & 14\% \\
$\bc{2}{2}{4}$ & 14 & 16 & 14\% \\
$\bc{3}{3}{3}$ & 23 & 26 & 17\% \\
$\bc{2}{3}{3}$ & 15 & 18 & 20\% \\
$\bc{2}{3}{4}$ & 20 & 24 & 20\% \\
$\bc{2}{4}{4}$ & 26 & 32 & 23\% \\
$\bc{3}{3}{4}$ & 29 & 36 & 24\% \\
$\bc{3}{4}{4}$ & 38 & 48 & 26\% \\
$\bc{3}{3}{6}$ \cite{smirnov2013bilinear} & 40 & 54 & 35\% \\
\midrule
$\bc{2}{2}{3}$* \cite{BCRL79} & 10 & 12 & 20\% \\
$\bc{3}{3}{3}$* \cite{Schonhage81} & 21 & 27 & 29\% \\
\bottomrule
\end{tabular}
\label{tab:algorithms}
\end{table}

\section{Implementation and practical considerations}

We now discuss our code generation method for fast algorithms and the major implementation issues.
All experiments were conducted on a single compute node on NERSC's Edison.
Each node has two 12-core Intel 2.4 GHz Ivy Bridge processors and 64 GB of memory.

\label{sec:implementation}

\subsection{Code generation}
\label{sec:codegen}

Our code generator automatically implements a fast algorithm in C++ given the $\M{U}$, $\M{V}$, and $\M{W}$ matrices representing the algorithm.
The generator simultaneously produces both sequential and parallel implementations.
We discuss the sequential code in this section and the parallel extensions in Section~\ref{sec:parallel}.
For computing $\M{C} = \M{A} \cdot \M{B}$, the following are the key ingredients of the generated code:
\begin{itemize}
\item
Using the entries in the $\M{U}$ and $\M{V}$ matrices, form the temporary matrices $\M{S}_r$ and $\M{T}_r$, $1 \le r \le R$, via matrix additions and scalar multiplication.
The $\M{S}_r$ and $\M{T}_r$ are linear combinations of sub-blocks of $\M{A}$ and $\M{B}$, respectively.
For each $\M{S}_r$ and $\M{T}_r$, the corresponding linear combination is customly generated.
Scalar multiplication by $\pm 1$ is replaced with native addition / subtraction operators.
The code generator can produce three variants of matrix additions, which we describe in Section~\ref{sec:matrix_additions}.

When a column of $\M{U}$ or $\M{V}$ contains a single non-zero element, there is no matrix addition (only scalar multiplication).
In order to save memory, the code generator does not form a temporary matrix in this case.
The scalar multiplication is piped through to subsequent recursive calls and is eventually used in a base case call to $\dgemm$.

\item
Recursive calls to the fast matrix multiplication routine compute $\M{M}_r = \M{S}_r \cdot \M{T}_r$, $1 \le r \le R$.

\item
Using the entries of $\M{W}$, linear combinations of the $\M{M}_r$ form the output $\M{C}$.
Matrix additions and scalar multiplications are again handled carefully, as above.

\item
Common subexpression elimination detects redundant matrix additions, and the code generator can automatically implement
algorithms with fewer additions.
We discuss this process in more detail in Section~\ref{sec:cse}.

\item
Dynamic peeling handles arbitrary matrix dimensions to make the implementation general.
We review this procedure in in Section~\ref{sec:all_dimensions}.

\end{itemize}

\begin{figure}[tb]
\centering
\includegraphics[width=2.5in]{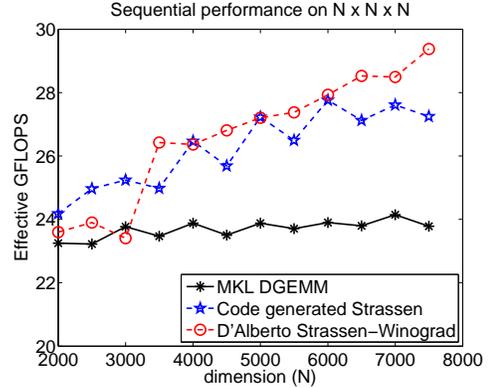}
\caption{
Effective performance (Equation~\eqref{eqn:eff_perf}) of our code generator's implementation of Strassen's algorithm against MKL's $\dgemm$ and a tuned implementation of the Strassen-Winograd algorithm \cite{d2011exploiting}.
The problems sizes are square.
The generated code easily outperforms MKL and is competitive with the tuned code.
}
\label{fig:comp_perf}
\end{figure}

Figure~\ref{fig:comp_perf} benchmarks the performance of the code generator's implementation.
In order to compare the performance of matrix multiplication algorithms with different computational costs,
we use the \emph{effective} GFLOPS metric for $\dims{P}{Q}{R}$ matrix multiplication:
\begin{equation}
\label{eqn:eff_perf}
\text{effective GFLOPS} = \frac{2PQR-PR}{\text{time in seconds}} \cdot 1\text{e-9}.
\end{equation}
We note that effective GFLOPS is only the true GFLOPS for the classical algorithm
(the fast algorithms perform fewer floating point operations).
However, this metric lets us compare all of the algorithms on an inverse-time scale, normalized by problem size \cite{lipshitz2012communication}.

We compare our code-generated Strassen implementation with MKL's $\dgemm$ and a tuned implementation of Strassen-Winograd from D'Alberto \emph{et al.} \cite{d2011exploiting} (recall that Strassen-Winograd performs the same number of multiplications but fewer matrix additions than Strassen's algorithm).
The code generator's implementation outperforms MKL and is competitive with the tuned implementation.
Thus, we are confident that the general conclusions we draw with code-generated implementations of fast algorithms will also apply to hand-tuned implementations.


\subsection{Handling matrix additions}
\label{sec:matrix_additions}

While the matrix multiplications constitute the bulk of the running time,
matrix additions are still an important performance optimization.
We call the linear combinations used to form $\M{S}_r$, $\M{T}_r$, and $\M{C}_{ij}$ \emph{addition chains}.
For example, $\M{S}_1 = \M{A}_{11} + \M{A}_{22}$ is an addition chain in Strassen's algorithm.
We consider three different implementations for the addition chains:

\begin{enumerate}
\item \textbf{Pairwise}:
With $r$ fixed, compute $\M{S}_r$ and $\M{T}_r$ using the $\daxpy$ BLAS routine for all matrices in the addition chain.
This requires $\nnz{\MC{U}{r}}$ calls to $\daxpy$ to form $\M{S}_r$ and $\nnz{\MC{V}{r}}$ calls to form $\M{T}_r$. 
After the $\M{M}_r$ matrices are computed recursively, we follow the same strategy to form the output.
The $i$th sub-block (row-wise) of $\M{C}$ requires $\nnz{\ME{W}{i,:}}$ $\daxpy$ calls.\footnote{Because $\daxpy$ computes $y\leftarrow \alpha x + y$, we make a call for each addition in the chain as well as one call for an initial copy.}

\item \textbf{Write-once}: With $r$ fixed, compute $\M{S}_r$ and $\M{T}_r$ with only one write for each entry (instead of, for example, $\nnz{\MC{V}{r}}$ writes for $\M{S}_r$ with the pairwise method).
In place of $\daxpy$, stream through the necessary submatrices of $\M{A}$ and $\M{B}$ and combine the entries to form $\M{S}_r$ and $\M{T}_r$.
This requires reading some submatrices of $\M{A}$ and $\M{B}$ several times,
but writing to only one output stream at a time.
Similarly, we write the output matrix $\M{C}$ once and read the $\M{M}_r$ several times.

\item \textbf{Streaming}: Read each input matrix once and write each temporary matrix $\M{S}_r$ and $\M{T}_r$ once.  Stream through the entries of each sub-block of $\M{A}$ and $\M{B}$, and update the corresponding entries in \emph{all} temporary matrices $\M{S}_r$ and $\M{T}_r$.
Similarly, stream through the entries of the $\M{M}_r$ and update \emph{all} submatrices of $\M{C}$.
\end{enumerate}

Each $\daxpy$ call requires two matrix reads and one matrix write (except for the first call in an addition chain, which is a copy and requires one read and one write).
Let $\nnz{\M{U}, \M{V}, \M{W}} = \nnz{\M{U}} + \nnz{\M{V}} + \nnz{\M{W}}$.
Then the pairwise additions perform $2 \cdot \nnz{\M{U}, \M{V}, \M{W}} - 2R - MN$ submatrix reads and $\nnz{\M{U}, \M{V}, \M{W}}$ submatrix writes.
However, the additions use an efficient vendor implementation.

The write-once additions perform $\nnz{\M{U}, \M{V}, \M{W}}$ submatrix reads and at most $2R + MN$ submatrix writes.
We do not need to write any data for the columns of $\M{U}$ and $\M{V}$ with a single non-zero entry.
These correspond to addition chains that are just a copy, for example, $\M{T}_{2} = \M{B}_{11}$ in Strassen's algorithm.
While we perform fewer reads and writes than the pairwise additions,
the complexity of our code increases (we have to write our own additions), and we can no longer use a tuned $\daxpy$ routine.
However, we do not worry about code complexity because we use code generation.
Since the problem is bandwidth-bound and compilers can automatically vectorize \texttt{for} loops, we don't expect the latter concern to be an issue.

Finally, the streaming additions perform $MK + KN + R$ submatrix reads and at most $2R + MN$ submatrix writes.
This is fewer reads than the write-once additions, but we have increased the complexity of the writes.
Specifically, we alternate writes to different memory locations, whereas with the write-once algorithm, we write to a single output stream.

The three methods also have different memory footprints.
With pairwise or write-once, $\M{S}_r$ and $\M{T}_r$ are formed just before computing $\M{M}_r$.
After $\M{M}_r$ is computed, the memory becomes available.
On the other hand, the streaming algorithm must compute all temporary matrices $\M{S}_r$ and $\M{T}_r$ simultaneously,
and hence needs $R$ times as much memory for the temporary matrices.
We will explore the performance of the three methods at the end of Section~\ref{sec:cse}.

\subsection{Common subexpression elimination}
\label{sec:cse}

The $\M{S}_r$, $\M{T}_r$, and $\M{M}_r$ matrices often share subexpressions.
For example, in our $\bc{4}{2}{4}$ fast algorithm (see Table~\ref{tab:algorithms}),
$\M{T}_{11}$ and $\M{T}_{25}$ are:
\begin{align*}
\M{T}_{11} &= \M{B}_{24} - \M{B}_{12} - \M{B}_{22} & \M{T}_{25} &= \M{B}_{23} + \M{B}_{12} + \M{B}_{22}
\end{align*}

Both $\M{T}_{11}$ and $\M{T}_{25}$ share the subexpression $\M{B}_{12} + \M{B}_{22}$, up to scalar multiplication.
Thus, there is opportunity to remove additions / subtractions:
\begin{align*}
\M{Y}_1 &= \M{B}_{12} + \M{B}_{22} & \M{T}_{11} &= \M{B}_{24} - \M{Y}_1 & \M{T}_{25} &= \M{B}_{23} + \M{Y}_1
\end{align*}

\begin{table}[tb]
\centering
\caption{
Number of additions saved by greedily eliminating length-two common subexpressions in the formation of
the $\M{S}$ and $\M{T}$ matrices.
Since a single subexpression may be used several times, the number of additions saved is greater than the
number of subexpressions eliminated.
}
\begin{tabular}{c c c c c}
\toprule
Algorithm   & Original & CSE & Subexpressions & Additions \\
base case  &              &                 & eliminated & saved \\ \midrule
$\bc{3}{3}{3}$   & 97 & 70 & 18 & 27\\
$\bc{4}{2}{4}$   & 189 & 138 & 25 & 51\\
$\bc{4}{3}{2}$   & 96 & 72 & 13 & 24\\
$\bc{4}{3}{3}$   & 164 & 125 & 26 & 39\\
$\bc{5}{2}{2}$   & 53 & 43 & 7 & 10 \\
\bottomrule
\end{tabular}
\label{tab:cse}
\end{table}

Table~\ref{tab:cse} shows how many additions are saved when greedily eliminating length-two expressions.
At face value, eliminating additions would appear to improve the algorithm.
However, there are two important considerations.
First, using $\M{Y}_1$ with the pairwise or write-once approaches requires additional memory (with the streaming approach it requires only additional local variables).

Second, we discussed in Section~\ref{sec:matrix_additions} that an important metric
is the number of reads and writes.
If we use the write-once algorithm, we have actually \emph{increased} the number of reads and writes.
Originally, forming $\M{T}_{11}$ and $\M{T}_{25}$ required six reads and two writes.
By eliminating the common subexpression, we performed two fewer reads in forming $\M{T}_{11}$ and $\M{T}_{25}$ but needed
an additional two reads and one write to form $\M{Y}_1$.
In other words, we have read the same amount of data and written \emph{more} data.
In general, eliminating the same length-two subexpression $k$ times reduces the number of matrix reads and writes by $k - 3$.
Thus, a length-two subexpression must appear at least four times for elimination to reduce the total number of reads and writes in the algorithm.

\begin{figure*}[tb]
\centering
\includegraphics[width=1.72in]{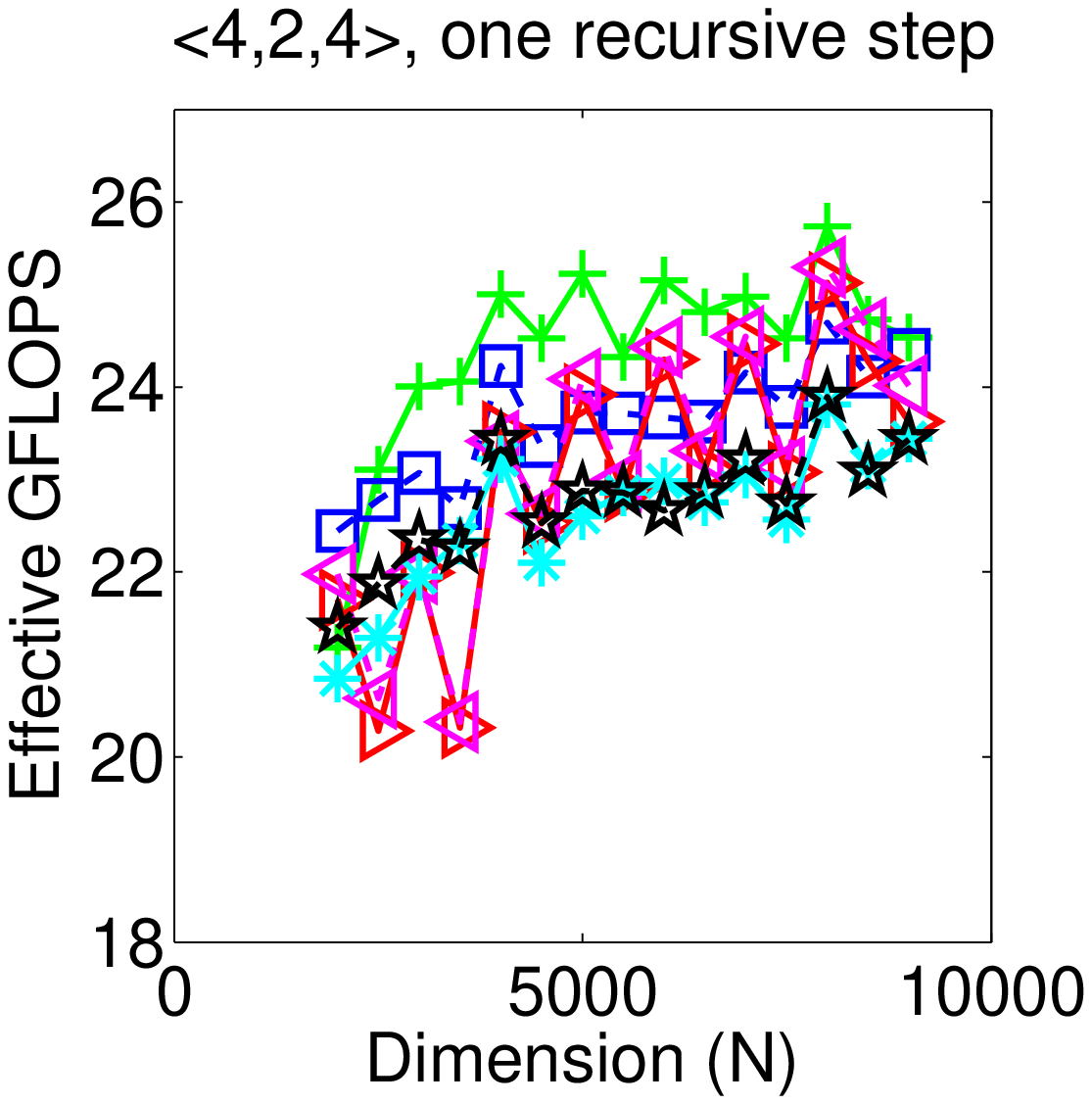}
\includegraphics[width=1.72in]{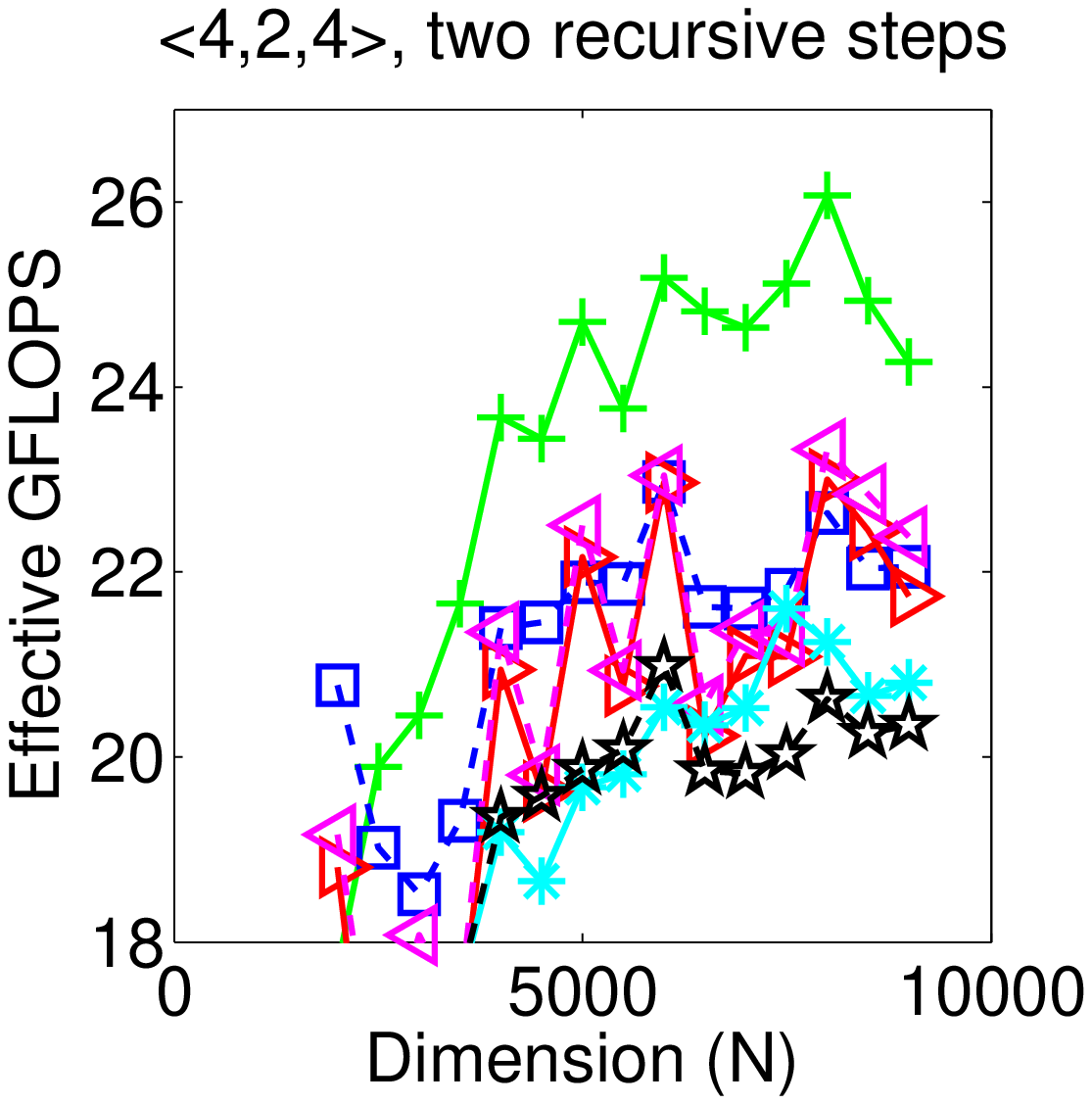}
\includegraphics[width=1.72in]{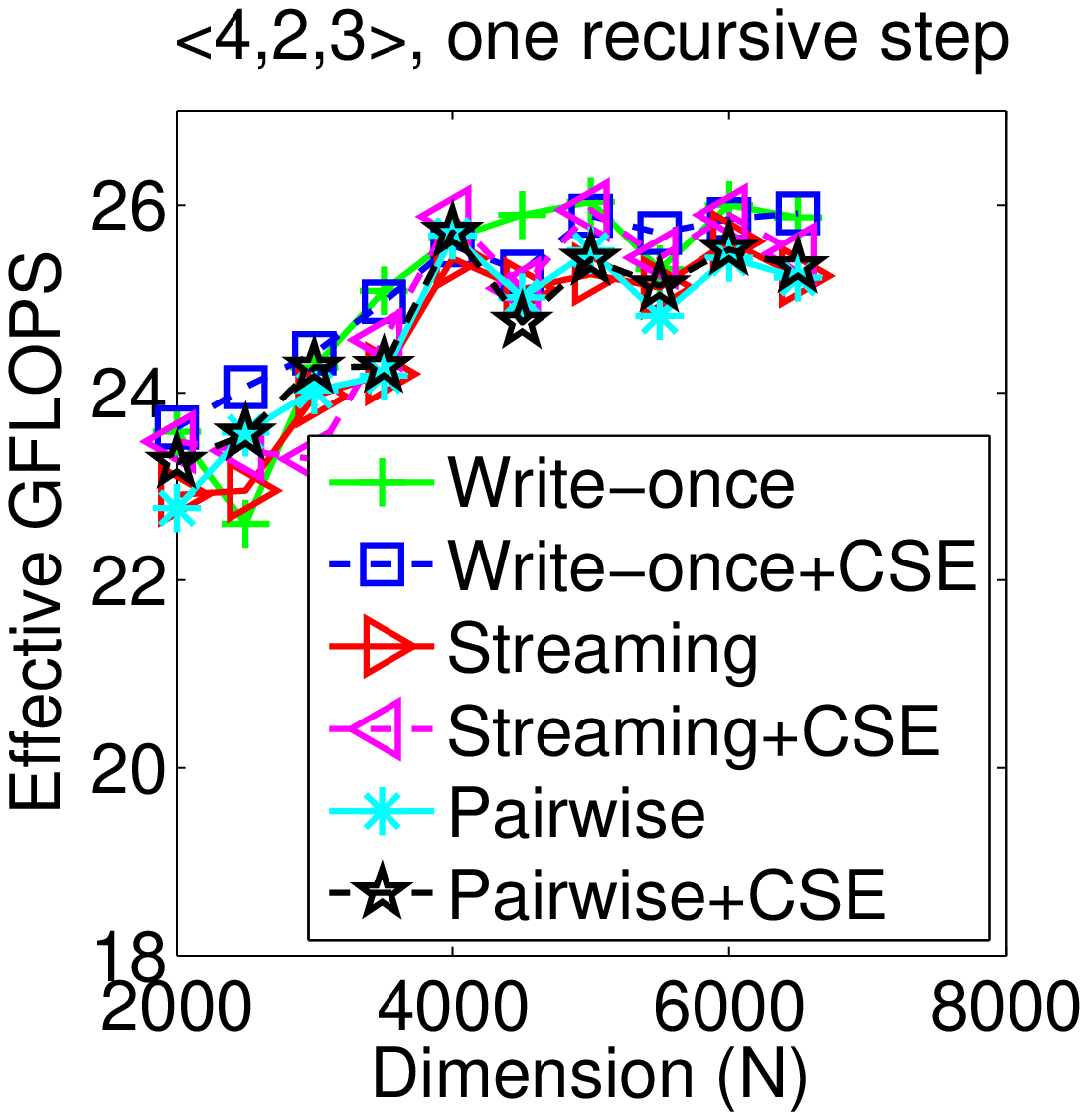}
\includegraphics[width=1.72in]{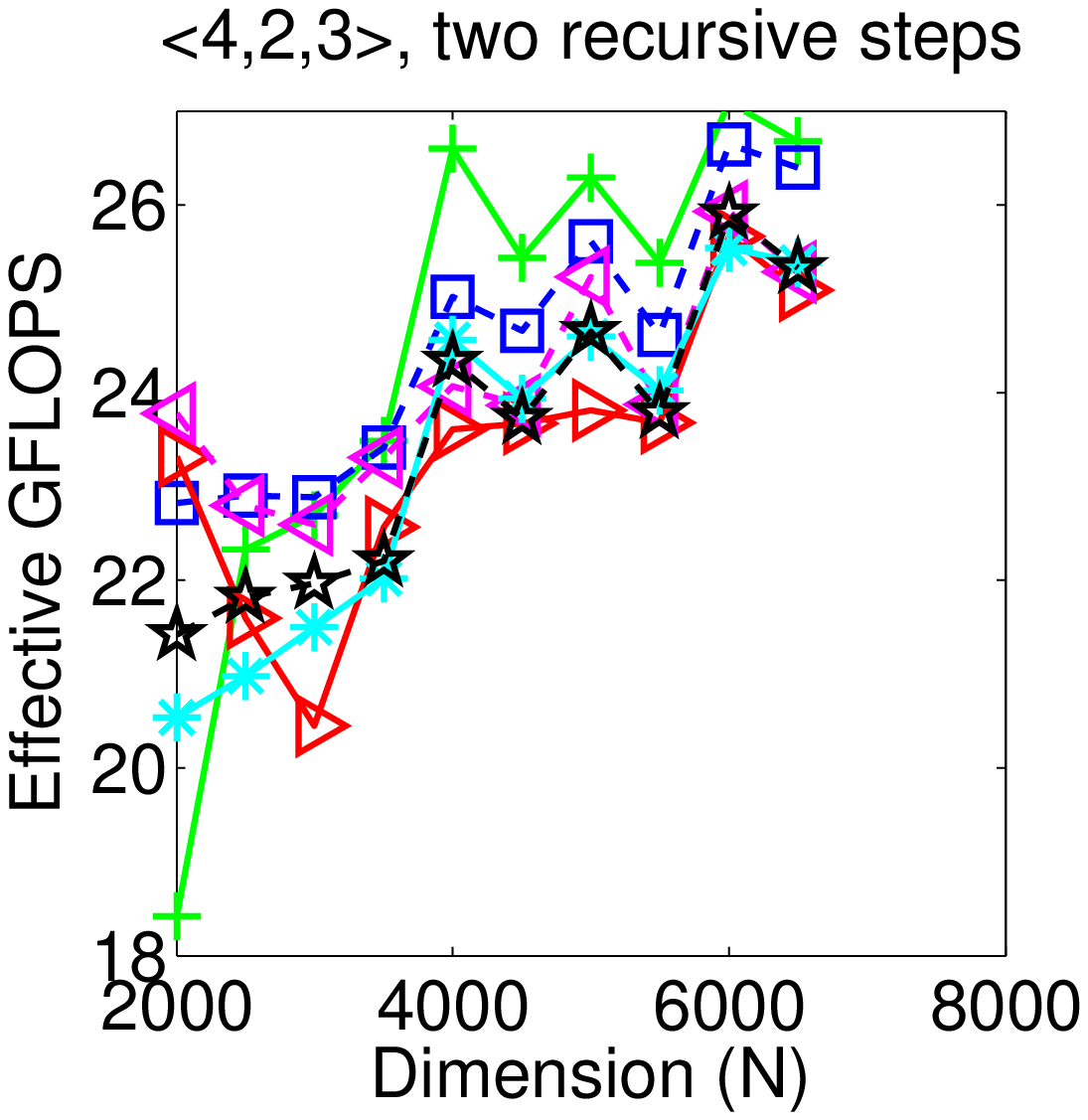}
\caption{
Effective performance (Equation~\eqref{eqn:eff_perf}) comparison of common subexpression elimination (CSE) and the three matrix addition methods: write-once, streaming, and pairwise (see Section~\ref{sec:matrix_additions}).
We use the code generator to implement six variants of fast algorithms for $\bc{4}{2}{4}$ and $\bc{4}{2}{3}$: using CSE or not for each of the three addition variants.
The $\bc{4}{2}{4}$ fast algorithm computed $\dims{N}{1600}{N}$ (``outer product" shape) for varying $N$,
and the $\bc{4}{2}{3}$ fast algorithm computed $\dims{N}{N}{N}$ (square multiplication).
For the $\bc{4}{2}{4}$ fast algorithm, no CSE with write-once additions has the highest performance;
for the $\bc{4}{2}{3}$ fast algorithm, it is less clear.
The pairwise variants tend to be slower because they perform more reads and writes.
}
\label{fig:adds_cse}
\end{figure*}

In Figure~\ref{fig:adds_cse}, we benchmark all three matrix addition algorithms from Section~\ref{sec:matrix_additions}, with and without common subexpression elimination.
In general, we see that the write-once algorithm without common subexpression elimination performs the best on the rectangular matrix multiplication problem sizes.
For these problems, common subexpression elimination lowers performance of the write-once algorithm and has little to modest effect on the streaming and pairwise algorithms.
For square matrix problems, the best variant is less clear, but write-once with no elimination often performs the highest.
We use write-once without elimination for the rest of our performance experiments.

\subsection{Recursion cutoff point}
\label{sec:recursion_cutoff}

In practice, we take only a few steps of recursion before calling a vendor-tuned library classical routine as the base case (in our case, Intel MKL's $\dgemm$).
One method for determining the cutoff point is to benchmark each algorithm and measure where the implementation
outperforms $\dgemm$.
While this is sustainable for the analysis of any individual algorithm, we are interested in a large class of fast algorithms.
Furthermore, a simple set of cutoff points limits understanding of the performance and will have to be re-measured for different architectures.
Instead, we provide a rule of thumb based on the performance of $\dgemm$.

\begin{figure*}[tb]
\centering
\includegraphics[width=3in]{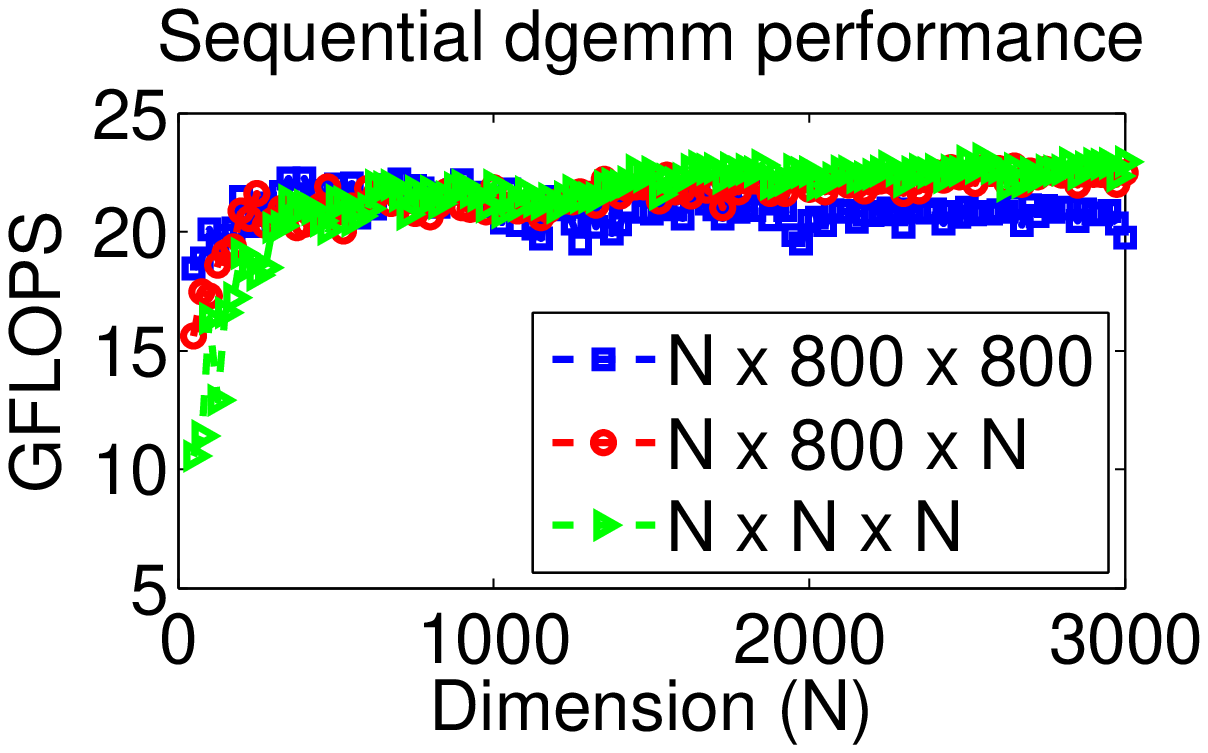} 
\includegraphics[width=3in]{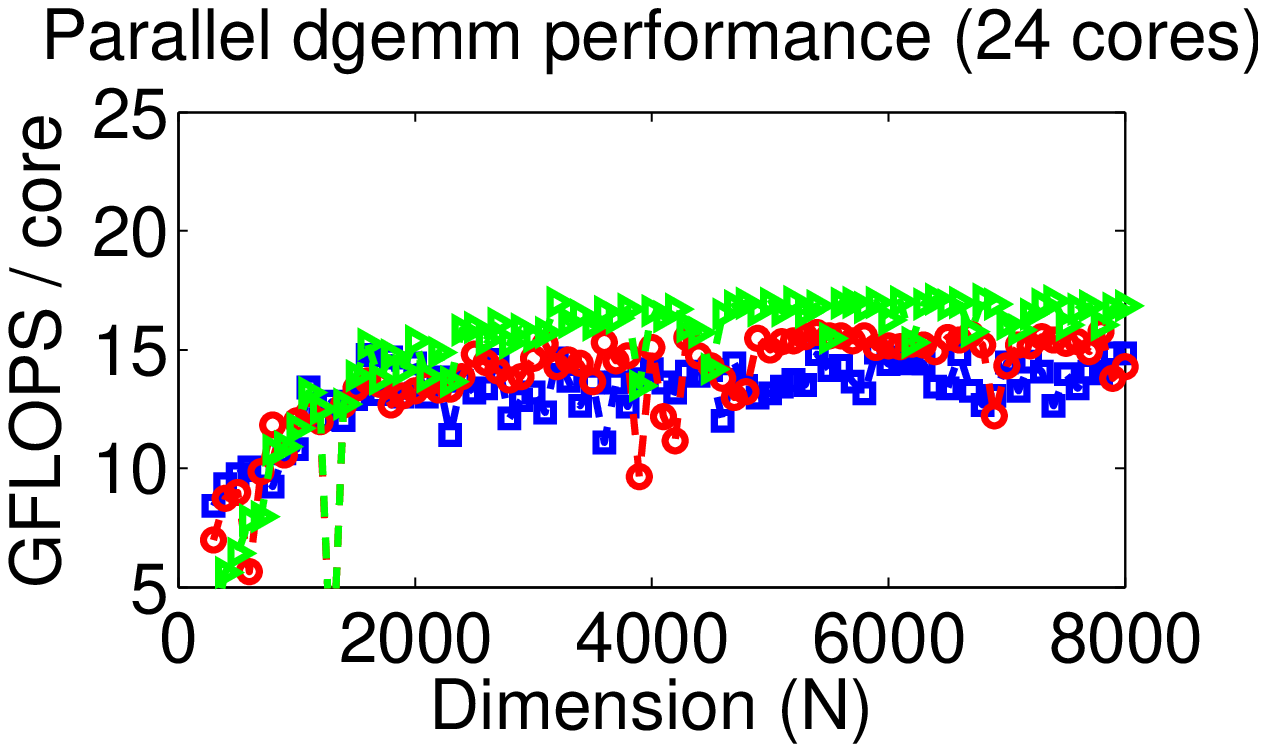}
\caption{
Performance curves of MKL's $\dgemm$ routine in serial (left) and in parallel (right) for three different problem shapes.
The performance curves exhibit a ``ramp-up" phase and then flatten for large enough problems.
Performance levels near $N = 1500$ in serial and $N = 5000$ in parallel.
For large problems in both serial and parallel, $\dims{N}{N}{N}$ multiplication is faster than $\dims{N}{800}{N}$, which is faster than $\dims{N}{800}{800}$.
We note that sequential performance is faster than parallel performance due to Intel Turbo Boost, which increases the clock speed from 2.4 to 3.2 GHz.
With Turbo Boost, peak sequential performance is 25.6 GFLOPS.
Peak parallel performance is 19.2 GFLOPS/core.
}
\label{fig:dgemm_curves}
\end{figure*}

Figure~\ref{fig:dgemm_curves} shows the performance of Intel MKL's sequential and parallel $\dgemm$ routines.
We see that the routines exhibit a ``ramp-up" phase and then flatten for sufficiently large problems.
In both serial and parallel, multiplication of square matrices ($\dims{N}{N}{N}$ computation) tends to level at a higher performance than
the problem shapes with a fixed dimension ($\dims{N}{800}{N}$ and $\dims{N}{800}{800}$).
Our principle for recursion is to take a recursive step only if the sub-problems fall on the flat part of the curve.
If the ratio of performance drop in the DGEMM curve is greater than the speedup per step (as listed in Table~\ref{tab:algorithms}),
then taking an additional recursive step cannot improve performance.\footnote{Note that the inverse is not necessarily true, the speedup depends on the overhead of the additions.}
Finally, we note that some of our parallel algorithms call the sequential $\dgemm$ routine in the base case.
Both curves will be important to our parallel fast matrix multiplication algorithms in Section~\ref{sec:parallel}.

\subsection{Handling arbitrary matrix dimensions}
\label{sec:all_dimensions}

The algorithms implemented by our code generator work for any matrix dimensions.
In order to use fast algorithms, however, the submatrices must be the same size---we have to add them together to form the $\M{S}_r$ and $\M{T}_r$ matrices.
There are several strategies for handling matrices whose dimensions are not evenly divided.
These include padding the matrix with zeroes, overlapping submatrices \cite{douglas1994gemmw}, and dynamic peeling \cite{thottethodi1998tuning}.
We choose dynamic peeling, which handles the boundaries of the matrix at each recursive level, in order to keep the code generation simple and limit memory consumption.

\section{Parallel algorithms for shared memory}
\label{sec:parallel}

We present three algorithms for parallel fast matrix multiplication: depth-first search (DFS), breadth-first search (BFS), and a hybrid of the two (HYBRID).
In this work, we target shared memory machines, although the same ideas generalize to distributed memory.
For example, DFS and BFS ideas are used for a distributed memory implementation of Strassen's algorithm~\cite{lipshitz2012communication}.

\subsection{Depth-first search}
\label{sec:par_dfs}

The DFS algorithm is straightforward: when recursion stops, the classical algorithm uses all threads on each sub-problem.
In other words, we use parallel matrix multiplication on the leaf nodes of a depth-first traversal of the recursion tree.
At a high-level, the code path is exactly the same as in the sequential case, and the main parallelism is in library calls.
The advantages of DFS are that the memory footprint matches the sequential algorithm and the code is simpler---parallelism in multiplications is hidden inside library calls.
Furthermore, matrix additions are trivially parallelized.
The key disadvantage of DFS is that the base case must be large enough to see a speed-up because the ramp-up curve is flatter (Figure~\ref{fig:dgemm_curves}).
For Edison's 24-core compute node, the base case should be around $N = 5000$.

\subsection{Breadth-first search}
\label{sec:par_bfs}

The BFS algorithm uses task-based parallelism.
Each leaf node in the matrix multiplication recursion tree is an independent task.
The recursion tree also serves as a dependency graph: we need to compute all $\M{M}_r$, $1 \le r \le R$, (children) before forming the result (parent).
The major advantage of BFS is we can take more recursive steps because the recursion cutoff point is based on the sequential $\dgemm$ curves.
Matrix additions to form $\M{S}_r$ and $\M{T}_r$ are part of the task to form the $\M{M}_r$.
In the first level of recursion, matrix additions to form $\M{C}_{ij}$ from the $\M{M}_r$ 
are handled in the same way as DFS, since all threads are available.

The BFS approach has two distinct disadvantages.
First, it is difficult to load balance the tasks because the number of threads may not divide the number of tasks evenly.
Also, with only one step of recursion, the number of tasks can be smaller than the number of threads.
For example, one step of Strassen's algorithm produces only 7 tasks and one step of the fast $\bc{3}{2}{3}$ algorithm produces only 15 tasks.
Second, BFS requires additional memory since the tasks are executed independently.
In a fast algorithm for $\bc{M}{K}{N}$ with $R$ multiplies, each recursive step requires a factor $R / (MN)$ more memory than the output matrix $\M{C}$ to store the $\M{M}_r$.
There are additional memory requirements for the $\M{S}_r$ and $\M{T}_r$ matrices, as discussed in Section~\ref{sec:matrix_additions}.

\subsection{Hybrid}
\label{sec:par_hybrid}

Our hybrid algorithm compensates for the load imbalance in BFS by applying the DFS approach on a subset of the base case problems.
With $L$ levels of recursion and $P$ threads, the hybrid algorithm applies task parallelism (BFS) to the first $R^L - (R^L \mod P)$ multiplications.
The number of BFS subproblems is a multiple of $P$, so this part of the algorithm is load balanced.
On the remaining $R^L \mod P$ sub-problems, all threads are used on each multiplication (DFS).

An alternative approach uses another level of hybridization: evenly assign as many as possible of the remaining $R^L \mod P$ multiplications to disjoint subsets of $P'<P$ threads (where $P'$ divides $P$), and then finish off the still-remaining multiplications with all $P$ threads.
This approach reduces the number of small multiplications assigned to all $P$ threads where perfect scaling is harder to achieve.
However, it leads to additional load balancing concerns in practice and requires a more complicated task scheduler.

\subsection{Implementation}
\label{sec:par_implementation}

The code generation from Section~\ref{sec:codegen} produces code that can compile to the DFS, BFS, or HYBRID parallel algorithms.
We use OpenMP to implement each algorithm.
The overview of the parallelization is:
\begin{itemize}
\item \textbf{BFS}: Each recursive matrix multiplication routine and the associated matrix additions are launched as an OpenMP task.
At each recursive level, the \texttt{taskwait} barrier ensures that all $\M{M}_r$ matrices are available to form the output matrix.
\item \textbf{DFS}: Each $\dgemm$ call uses all threads.
Matrix additions are always fully parallelized.
\item \textbf{HYBRID}:
Matrix multiplies are either launched as an OpenMP task (BFS), or the number of MKL threads is adjusted for a parallel $\dgemm$ (DFS).
This is implemented with the \texttt{if} conditional clause of OpenMP tasks.
Again, \texttt{taskwait} barriers ensure that $\M{M}_r$ matrices are computed to form the output matrix.
We use an explicit synchronization scheme with OpenMP locks to ensure that the DFS steps occur \emph{after} the BFS tasks complete.
This ensures that there is no oversubscription of threads.
\end{itemize}

\subsection{Shared-memory bandwidth limitations}
\label{sec:bandwidth}

The performance gains of the fast algorithms rely on the cost of matrix multiplications to be much larger than the cost of matrix additions.
Since matrix multiplication is compute-bound and matrix addition is bandwidth-bound, these computations scale differently with the amount of parallelism.
For large enough matrices, MKL's $\dgemm$ achieves near-peak performance of the node (Figure~\ref{fig:dgemm_curves}).
On the other hand, the STREAM benchmark \cite{mccalpin1995survey} shows that the node achieves around a five-fold speedup in bandwidth with 24 cores.
In other words, in parallel, matrix multiplication is near 100\% parallel efficiency and matrix addition is near 20\% parallel efficiency.
The bandwidth bottleneck makes it more difficult for parallel fast algorithms to be competitive with parallel MKL.
To illuminate this issue, we will present performance results with both 6 and 24 cores.
Using 6 cores avoids the bandwidth bottleneck and leads to much better performance per core.

\subsection{Performance comparisons}

\begin{figure*}[tb]
\centering
\includegraphics[width=2.3in]{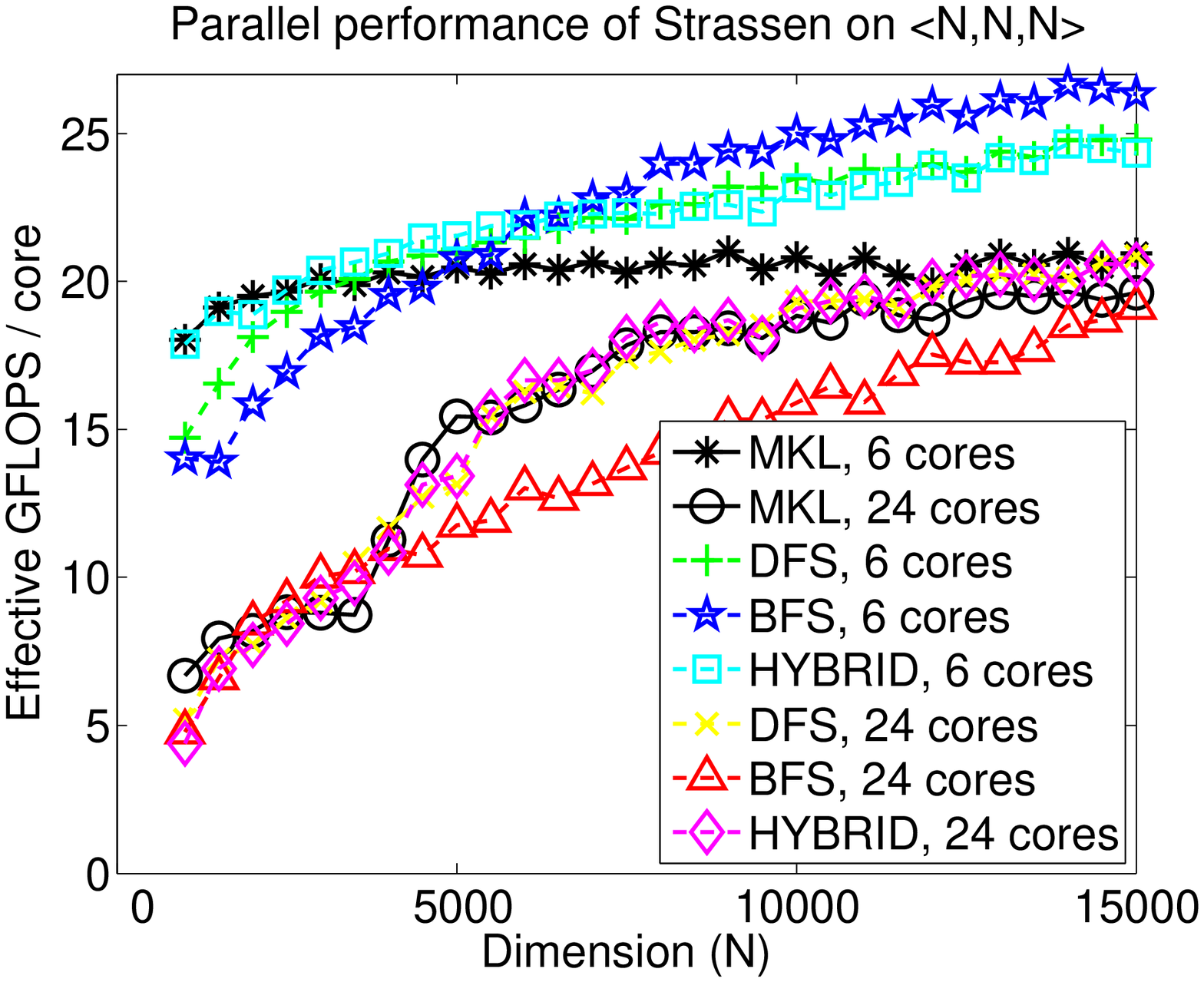}
\includegraphics[width=2.3in]{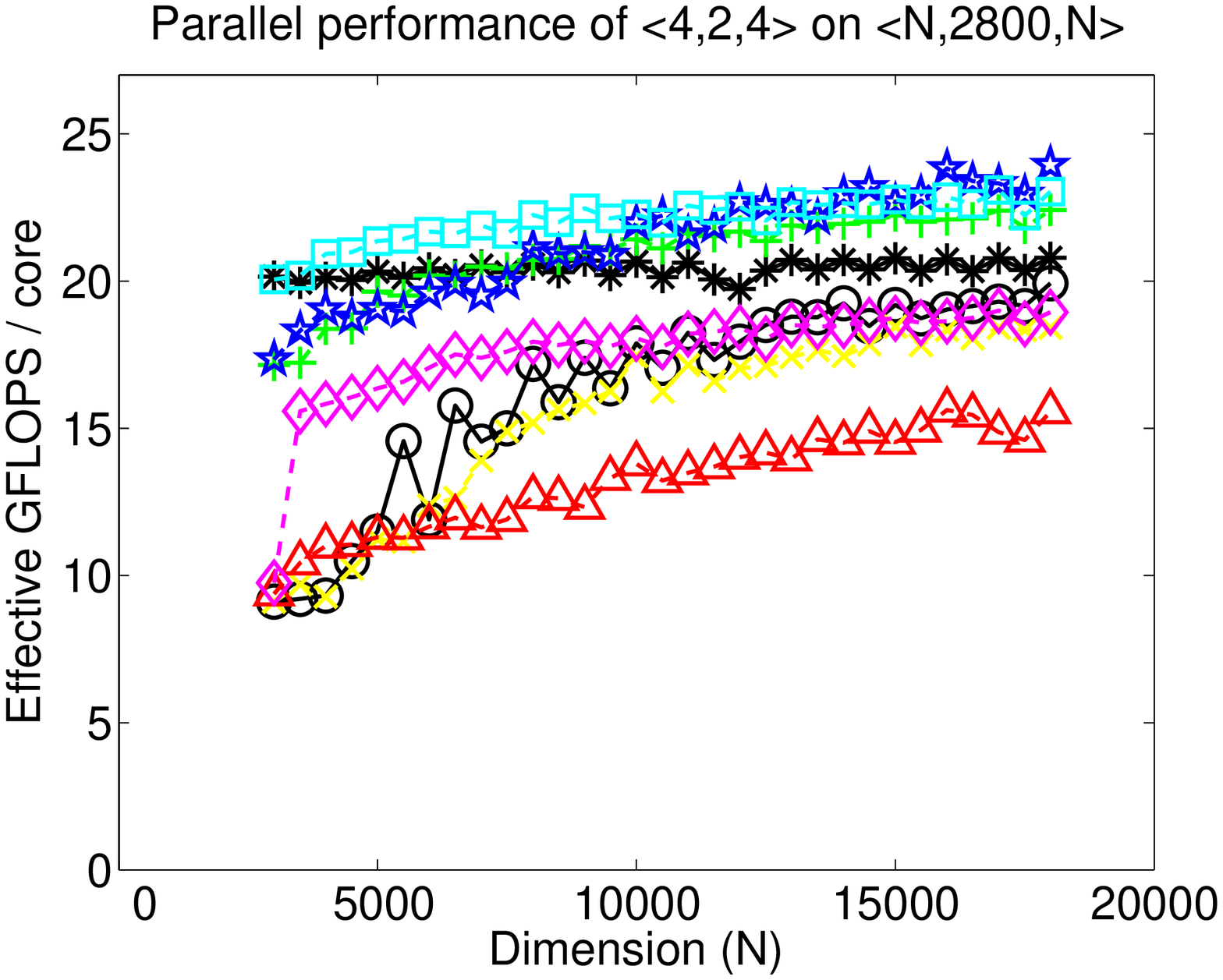}
\includegraphics[width=2.3in]{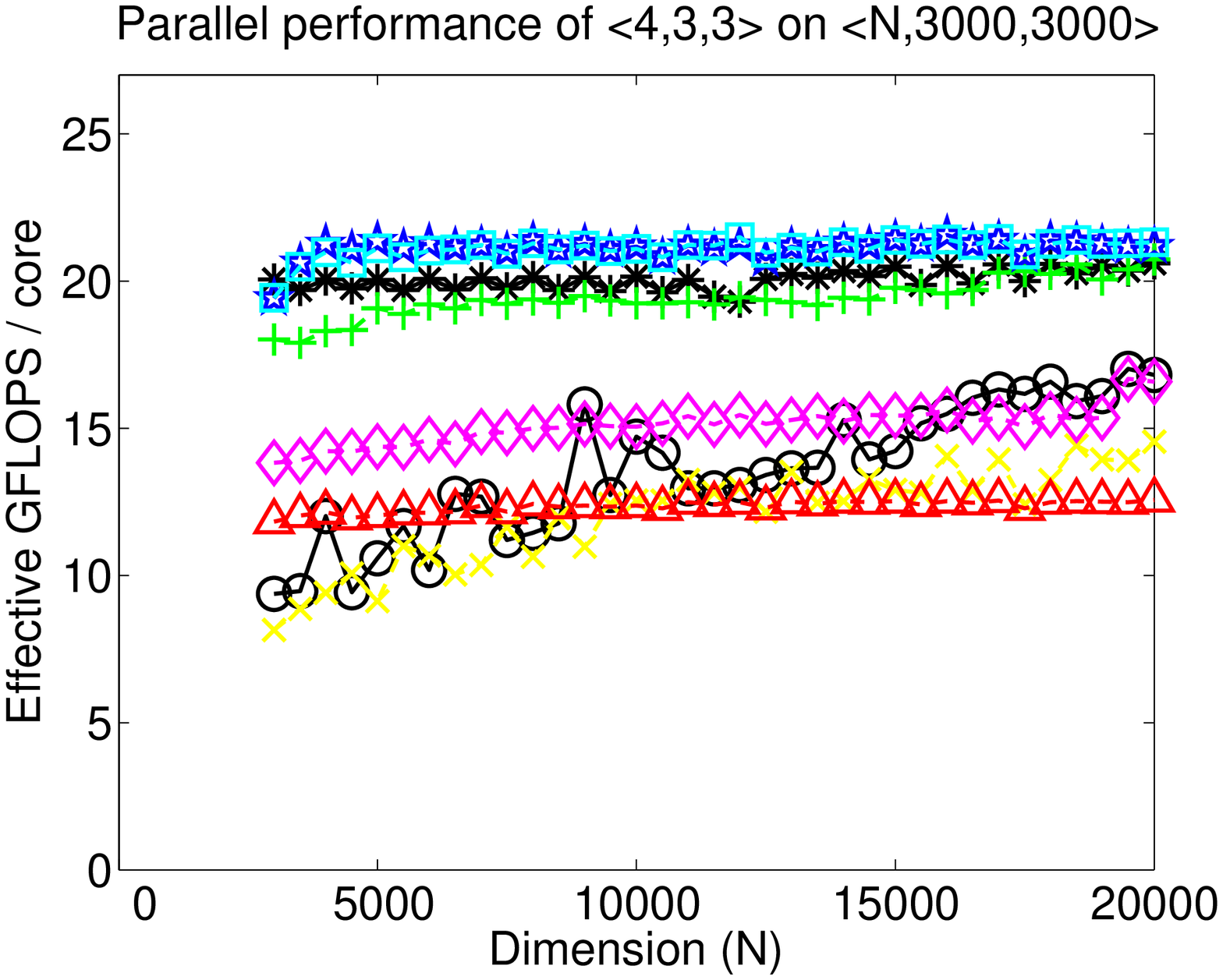}
\caption{
Effective performance (Equation~\eqref{eqn:eff_perf}) comparison of the BFS, DFS, and HYBRID parallel implementations on a few fast algorithms and problem sizes.
We use 6 and 24 cores to show the bandwidth limitations of the matrix additions.
(Left): 
Strassen's algorithm on square problems.
With 6 cores, we see significant speedups on large problems.
(Middle):
The $\bc{4}{2}{4}$ fast algorithm (26 multiplies) on $\dims{N}{2800}{N}$ problems.
HYBRID performs the best in all cases.
With 6 cores, the fast algorithm consistently outperforms MKL.
With 24 cores, the fast algorithm can achieve significant speedups for small problem sizes.
(Right):
The $\bc{4}{3}{3}$ fast algorithm (29 multiplies) on $\dims{N}{3000}{3000}$ problems.
HYBRID again performs the best.
With 24 cores, the fast algorithm gets modest speedups over MKL and achieves significant speedups on small problems.
}
\label{fig:comparisons}
\end{figure*}

Figure~\ref{fig:comparisons} shows the performance of the BFS, DFS, and HYBRID parallel methods with both 6 and 24 cores for three  representative algorithms.

The left plot shows the performance of Strassen's algorithm on square problems.
With 6 cores, HYBRID does the best for small problems.
Since Strassen's algorithm uses 7 multiplies, BFS has poor performance with 6 cores when using one step of recursion.
While all 6 cores can do 6 multiplies in parallel, the 7th multiply is done sequentially (with HYBRID, the 7th multiply uses all 6 cores).
With two steps of recursion, BFS has better load balance but is forced to work on smaller subproblems.
As the problems get larger, BFS outperforms HYBRID due to synchronization overhead when HYBRID switches from BFS to DFS steps.
When the matrix dimension is around 15,000, the fast algorithm achieves a 25\% speedup over MKL.
Using 24 cores, HYBRID and DFS are the fastest.
With one step of recursion, BFS can achieve only seven-fold parallelism.
With two steps, there are 49 subproblems, so one core is assigned 3 subproblems while all others are assigned 2.
In general, we see that it is much more difficult to achieve speedups with 24 cores.
However, Strassen's algorithm has a modest performance gain over MKL for large problem sizes ($\sim$ 5\% faster).

The middle plot of Figure~\ref{fig:comparisons} shows the $\bc{4}{2}{4}$ fast algorithm (26 multiplies) for $\dims{N}{2800}{N}$ problems.
With 6 cores, HYBRID is fastest for small problems and BFS becomes competitive for larger problems,
where the performance is 15\% better than MKL.
In Section~\ref{sec:performance}, we show that $\bc{4}{2}{4}$ is also faster than Strassen's algorithm for these problems.
With 24 cores, we see that HYBRID is drastically faster than MKL on small problems.
For example, HYBRID achieves a 75\% speedup on $\dims{3500}{2800}{3500}$. \footnote{
This result is an artifact of MKL's parallelization on these problem sizes and is not due to the speedups of the fast algorithm.
We achieved similar speedups using our code generator and a classical, $\bc{2}{3}{4}$ recursive algorithm (24 multiplies).
}
As the problem sizes get larger, we experience the bandwidth bottleneck and HYBRID achieves around the same performance as MKL.
BFS uses one step of recursion and is consistently slower since it parallelizes 24 of 26 multiplies and uses only 2 cores on the last 2 multiplies.
While multiple steps of recursion creates more load balance, the subproblems are small enough that performance degrades even more.
DFS follows a similar ramp-up curve as MKL, but the subproblems are still too small to see a performance benefit.

The right plot shows the $\bc{4}{3}{3}$ fast algorithm (29 multiplies) for $\dims{N}{3000}{3000}$.
We see similar trends as for the other problem sizes.
With 6 cores, HYBRID does well for all problem sizes.
Speedups are around $\sim$ 5\% for large problems.
With 24 cores, HYBRID is again drastically faster than MKL for small problem sizes and about the same as MKL for large problems.

\section{Performance experiments}
\label{sec:performance}

We now present performance results for a variety of fast algorithms on several problem sizes.
Based on the results of Section~\ref{sec:bandwidth},
we take the best of BFS and HYBRID when using 6 cores and the best of DFS and HYBRID when using 24 cores.
For rectangular problem sizes in both sequential and parallel, we take the best of one or two steps of recursion.
And for square problem sizes, we take the best of one, two, or three steps of recursion.
Additional recursive steps do not improve the performance for the problem sizes we consider.
The square problem sizes for parallel benchmarks require the most memory---for some algorithms, three steps of recursion results in out-of-memory errors.
In these cases, the original problem consumes 6\% of the memory.
For these algorithms, we only record the best of one or two steps of recursion in the performance plots.
Finally, all timings are the median of five trials.

\subsection{Sequential performance}
\label{sec:perf_sequential}

\begin{figure*}[tb]
\centering
\includegraphics[width=2.30in]{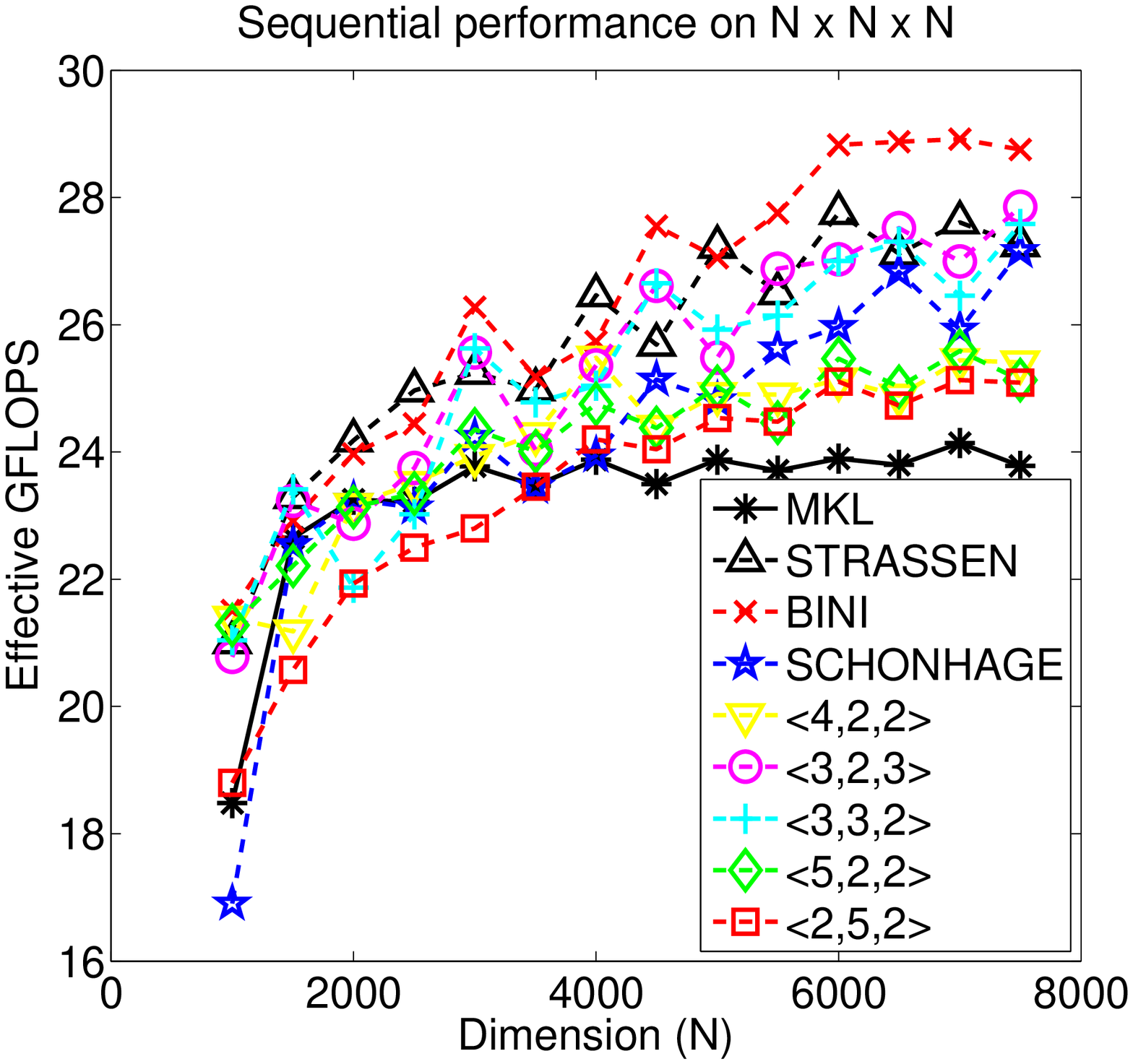}
\includegraphics[width=2.30in]{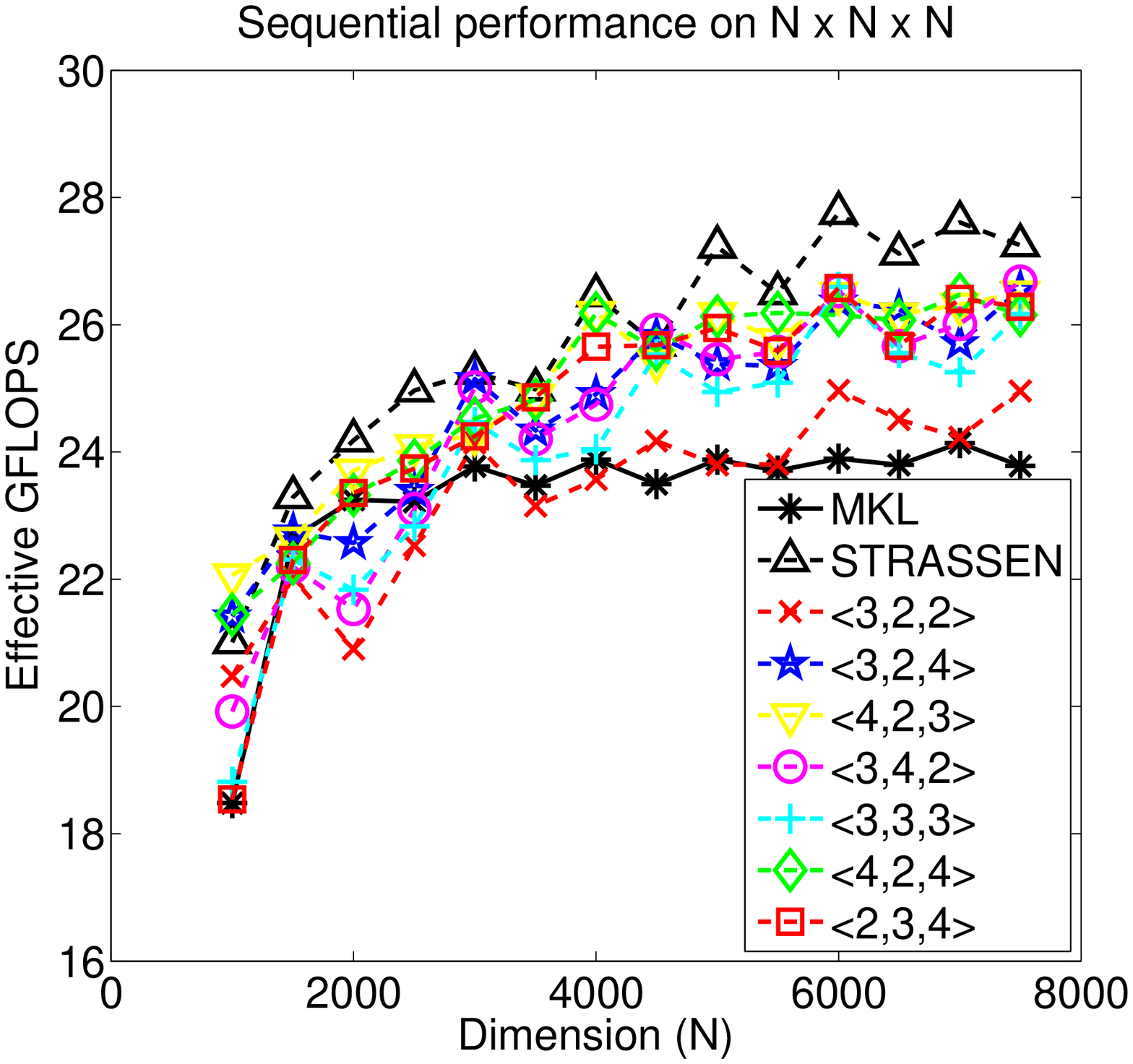}
\includegraphics[width=2.30in]{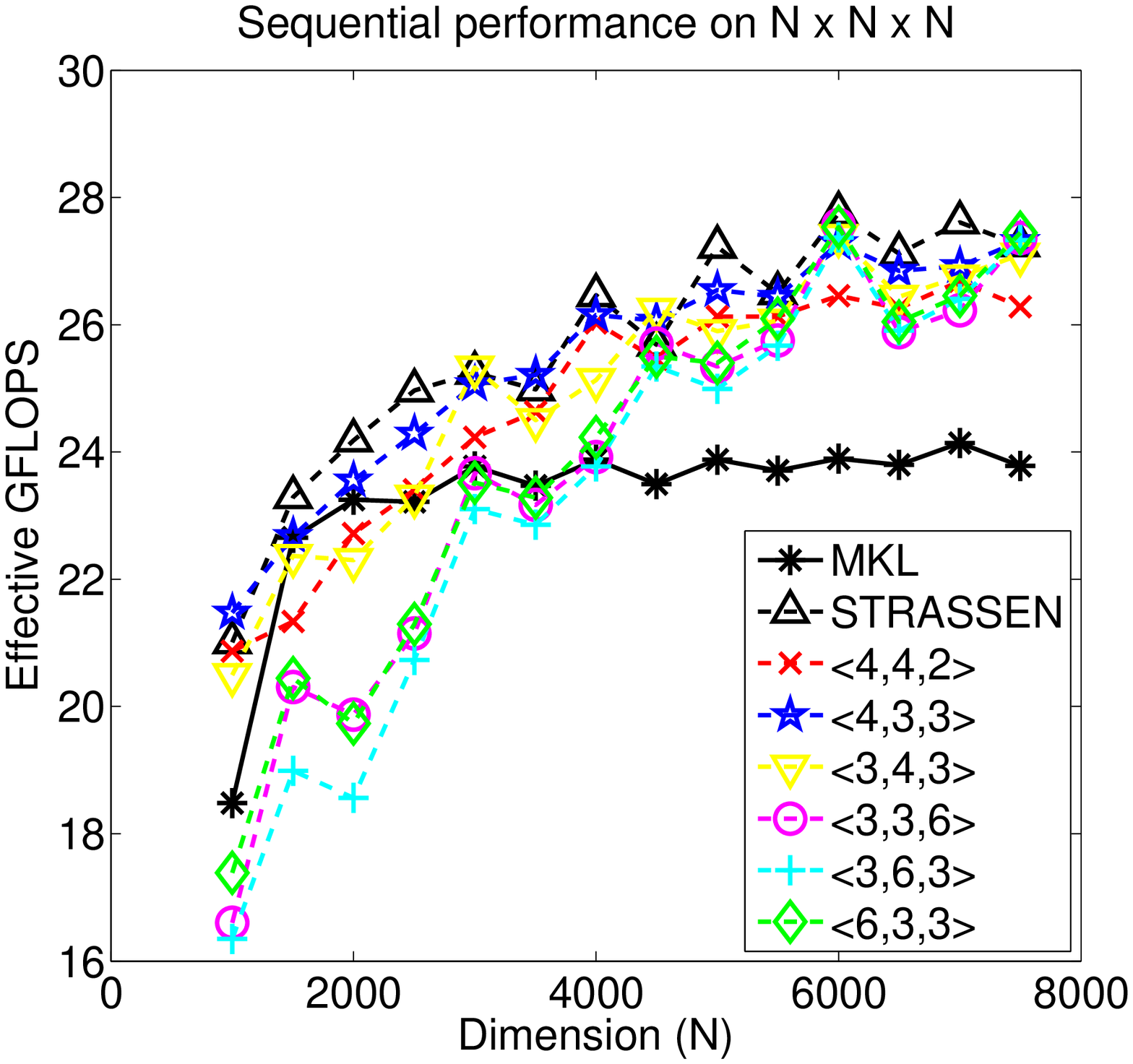} \\
\includegraphics[width=2.30in]{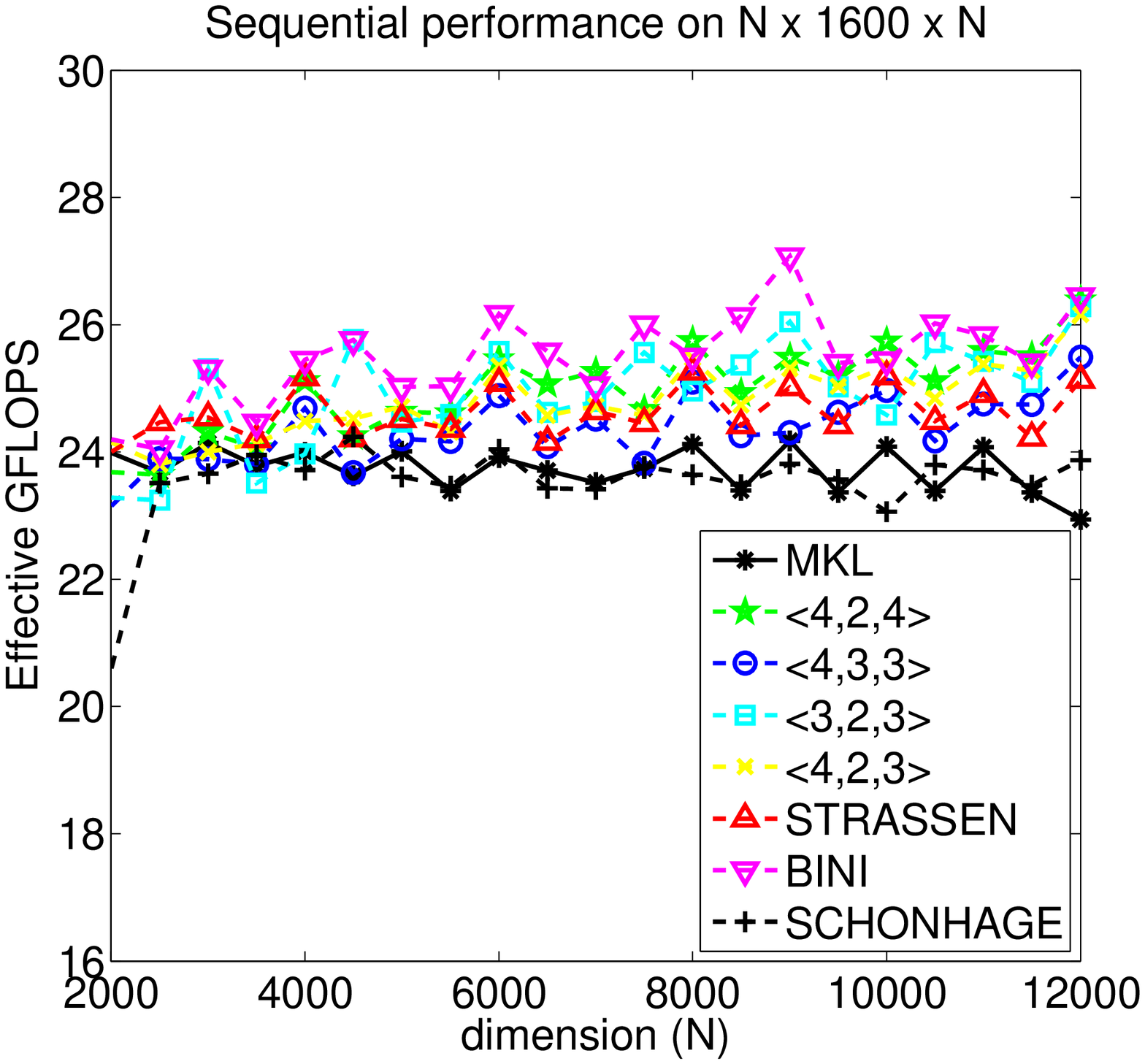}
\includegraphics[width=2.30in]{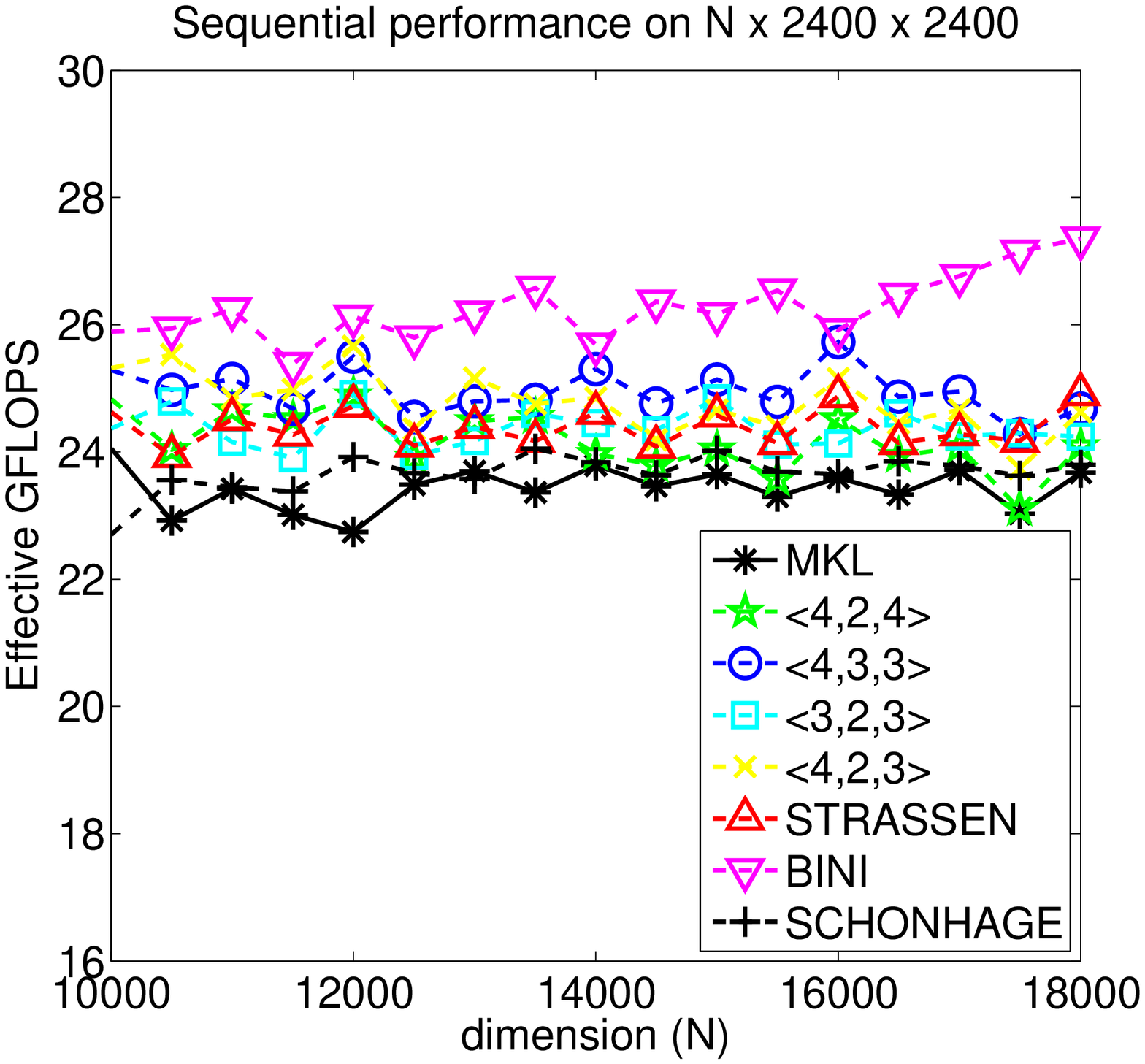}
\caption{
Effective sequential performance (Equation~\eqref{eqn:eff_perf}) of fast matrix multiplication algorithms.
Each data point is the best of one, two, or three steps of recursion: additional recursive steps did not improve performance.
MKL is a call to $\dgemm$,
Bini and Sch\"{o}nhage are approximate algorithms, 
and all others are exact fast algorithms.
(Top row):
Performance of a variety of fast algorithms on $\dims{N}{N}{N}$ problem sizes distributed across three plots.
MKL and Strassen are repeated on all three plots for comparison.
All of the fast algorithms outperform MKL for large enough problem sizes, and Strassen's algorithm usually performs the best.
(Bottom left):
Performance on an ``outer product" shape, $\dims{N}{1600}{N}$.
Exact fast algorithms that have a similar outer product shape (\emph{e.g.}, $\bc{4}{2}{4}$) tend to have the highest performance.
(Bottom right):
Performance of multiplication of tall-and-skinny matrix by a small square matrix, $\dims{N}{2400}{2400}$.
Again, fast algorithms that have this shape (\emph{e.g.}, $\bc{4}{3}{3}$) tend to have the highest performance.
}
\label{fig:sequential_perf}
\end{figure*}

Figure~\ref{fig:sequential_perf} summarizes the sequential performance of several fast algorithms.
For $\dims{N}{N}{N}$ problems, we test the algorithms in Table~\ref{tab:algorithms} and some of their permutations (top row of plots in Figure~\ref{fig:sequential_perf}).
For example, we test $\bc{4}{4}{2}$ and $\bc{4}{2}{4}$, which are permutations of $\bc{2}{4}{4}$.
In total, over 20 algorithms are tested for square matrices.
Two of these algorithms, Bini's $\bc{3}{2}{2}$ and Sch\"{o}nhage's $\bc{3}{3}{3}$ are APA algorithms.
We note that APA algorithms are of limited practical interest; even one step of recursion causes numerical errors in at least half the digits (a better speedup with the same or better numerical accuracy can be obtained by switching to single precision).
For the problem sizes $\dims{N}{1600}{N}$ and $\dims{N}{2400}{2400}$, we evaluate the APA algorithms and list performance for algorithms that are comparable to, or outperform, Strassen's algorithm.
The results are summarized as follows:

\begin{enumerate}
\item
All of the fast algorithms outperform MKL for large enough problem sizes.
These algorithms are implemented with our code generator and use only the high-level optimizations described in Section~\ref{sec:codegen}.
Since the fast algorithms perform less computation and communication, we expect this to happen.

\item
For square matrices, Strassen's algorithm often performs the best.
This is mostly due to its relatively small number of matrix additions in comparison to other fast algorithms.
On large problem sizes, Strassen's algorithm provides around a 20\% speedup over MKL's $\dgemm$.
In the top right plot of Figure~\ref{fig:sequential_perf}, we see that some algorithms become competitive with Strassen's algorithm for larger problem sizes.
These algorithms tend to have large speedups per recursive step (see Table~\ref{tab:algorithms}).
While Strassen's algorithm can take more recursive steps, memory constraints and the cost of additions with additional recursive steps cause Strassen's algorithm to be on par with these other algorithms.

\item\label{itm:nonsquare}
Although Strassen's algorithm has the highest performance for square matrices, other fast algorithms have higher performance for $\dims{N}{1600}{N}$ and $\dims{N}{2400}{2400}$ problem sizes (bottom row of Figure~\ref{fig:sequential_perf}).
The reason is that the fixed dimension constrains the number of recursive steps that can be taken by the fast algorithms.
With multiple recursive steps, the matrix sub-blocks become small enough so that $\dgemm$ does not achieve good performance on the subproblem.
Thus, fast algorithms that get a better speedup per recursive step typically have higher performance for these problem sizes.

\item
For rectangular matrices, algorithms that ``match the shape" of the problem tend to perform the best.
For example, $\bc{4}{2}{4}$ and $\bc{3}{2}{3}$ both have the ``outer product" shape of the $\dims{N}{1600}{N}$ problem sizes and have the highest performance.
Similarly, $\bc{4}{2}{3}$ and $\bc{4}{3}{3}$ have the highest performance of the exact algorithms for $\dims{N}{2400}{2400}$ problem sizes.
The  $\bc{4}{2}{4}$ and $\bc{4}{3}{3}$ algorithms provide around a 5\% performance improvement over Strassen and a 10\% performance improvement over MKL on $\dims{N}{1600}{N}$ and $\dims{N}{2400}{2400}$, respectively.
The reason follows from the performance explanation from Result~\ref{itm:nonsquare}.
Only one or two steps of recursion improve performance.
Algorithms that match the problem shape land have high speedups per recursive step perform the best.

\item
Bini's $\bc{3}{2}{2}$ APA algorithm typically has the highest performance on rectangular problem sizes.
However, we remind the reader that the approximation used by this algorithm results in severe numerical errors.

\end{enumerate}

\subsection{Parallel performance}
\label{sec:perf_parallel}

\begin{figure*}[tb]
\centering
\includegraphics[width=2.30in]{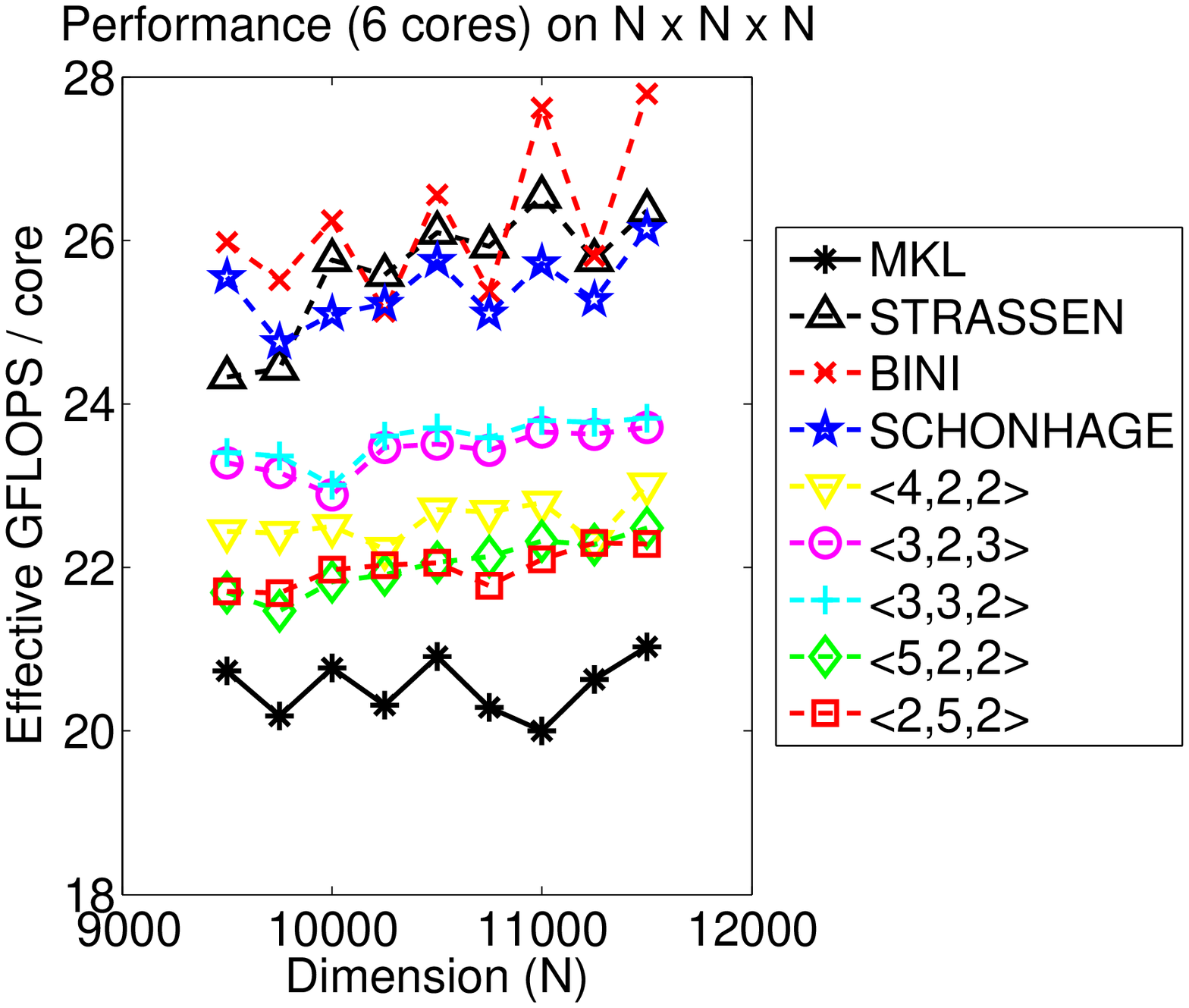}
\includegraphics[width=2.30in]{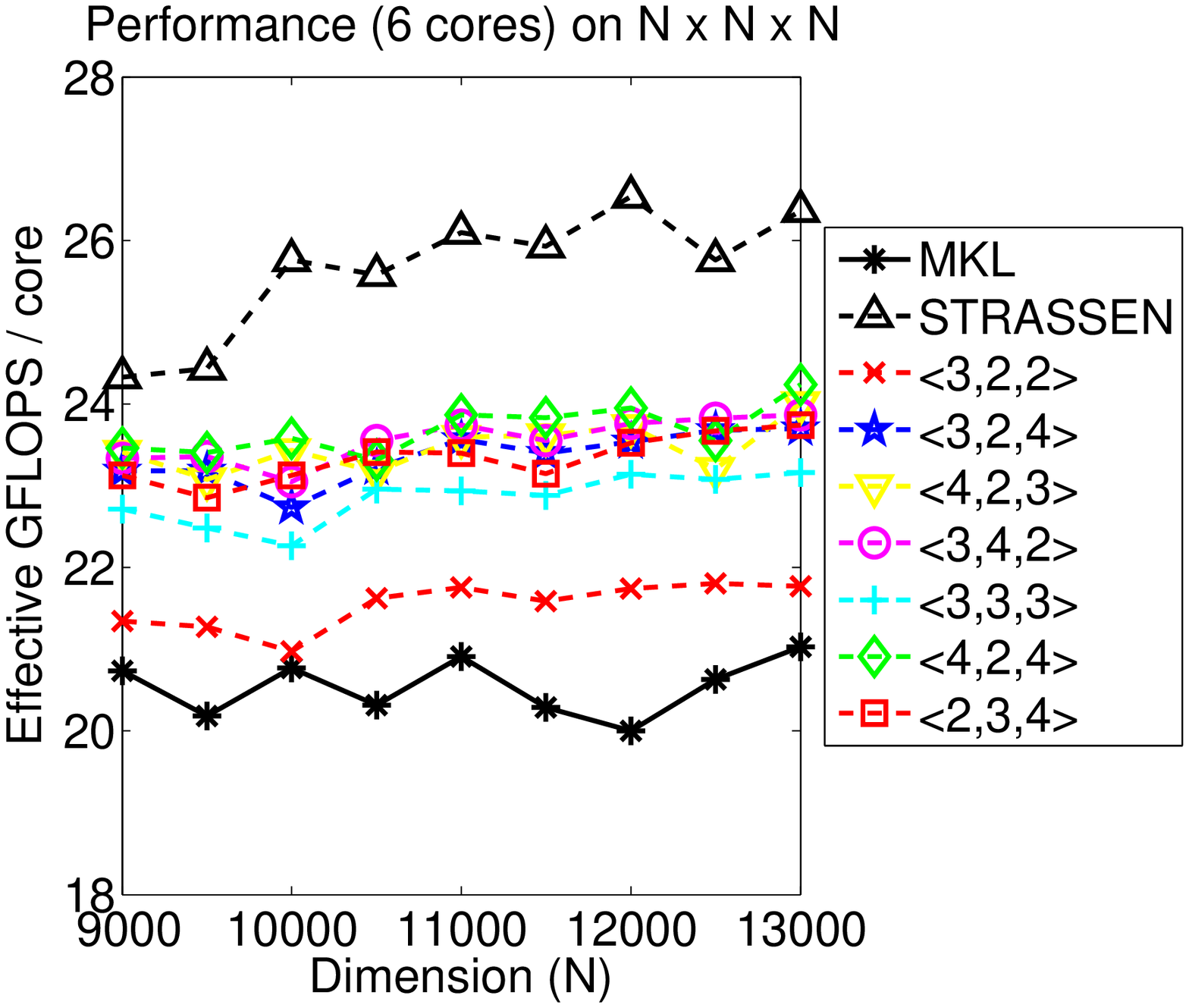}
\includegraphics[width=2.30in]{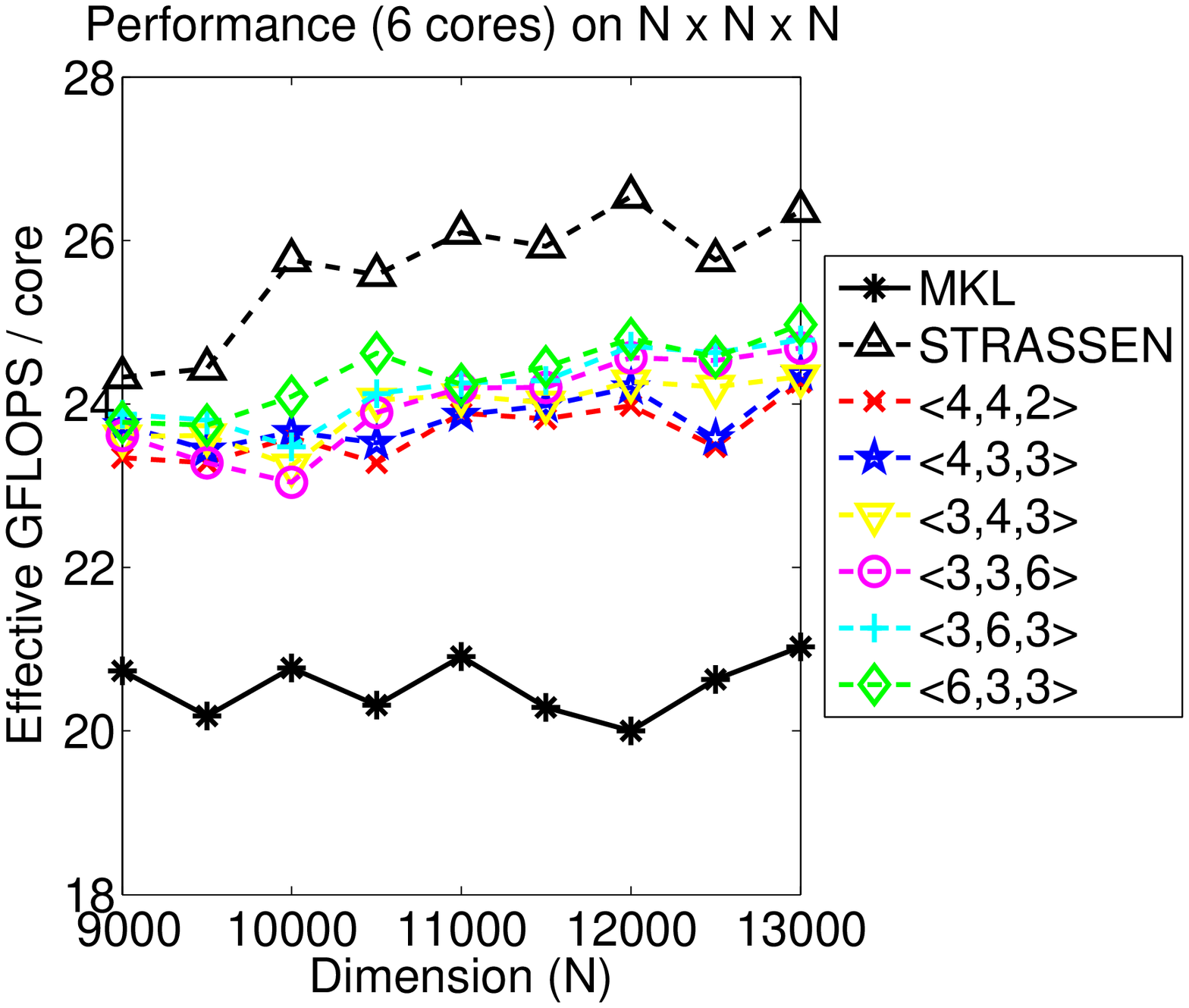} \\
\includegraphics[width=2.30in]{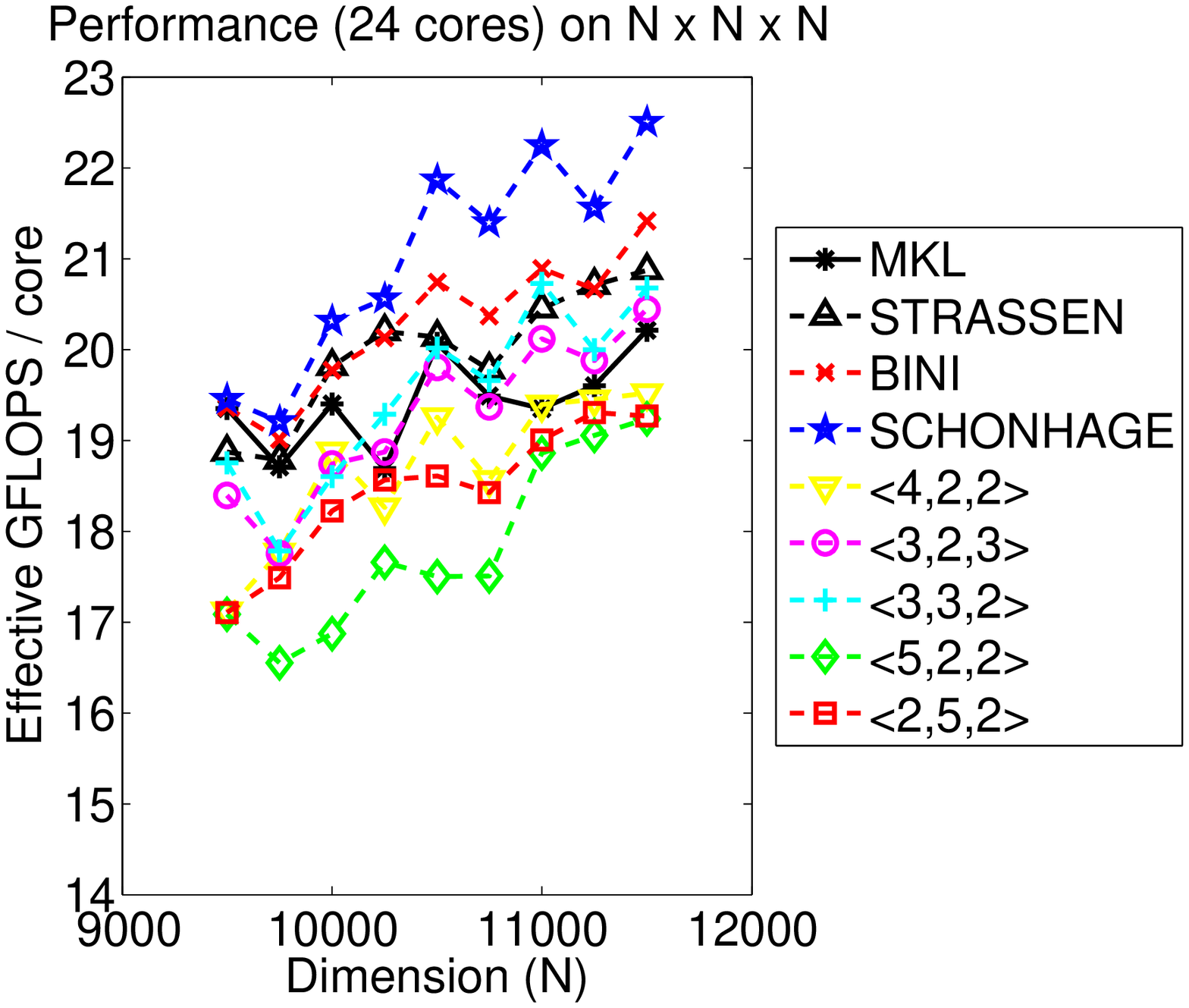}
\includegraphics[width=2.30in]{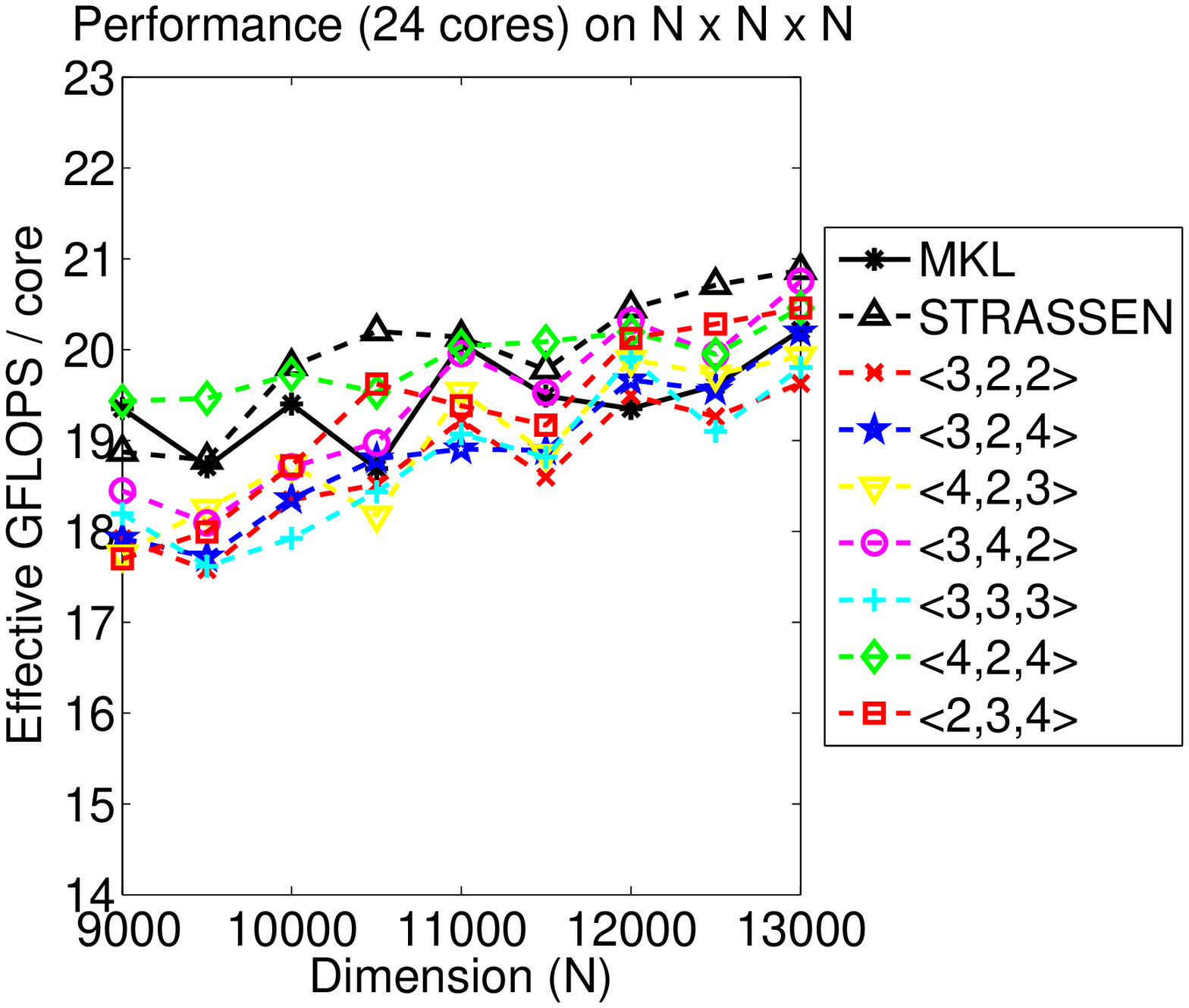}
\includegraphics[width=2.30in]{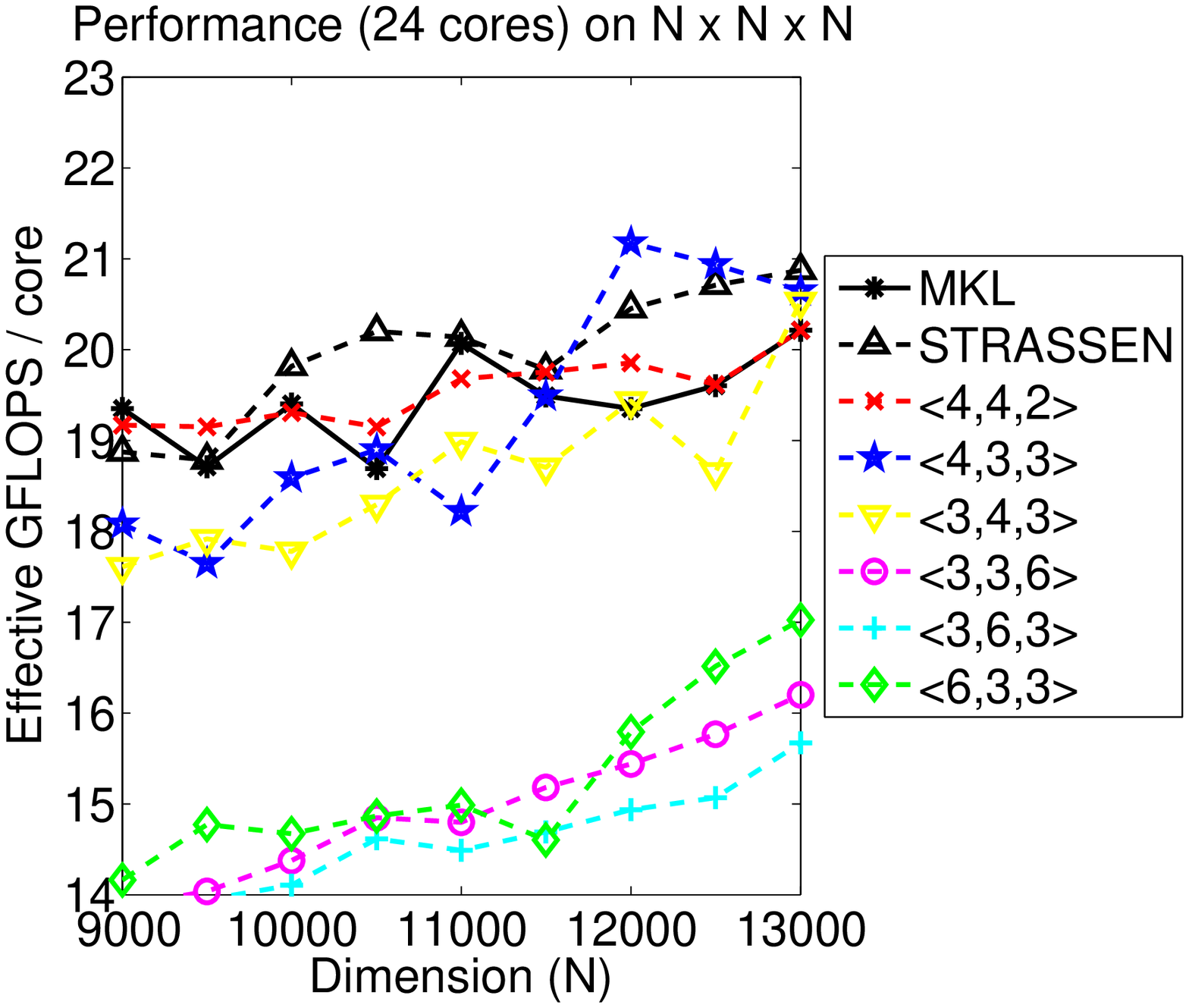}
\caption{
Effective parallel performance (Equation~\eqref{eqn:eff_perf}) of fast algorithms on square problems using only 6 cores (top row) and all 24 cores (bottom row).
With 6 cores, bandwidth is not a bottleneck and we see similar trends to the sequential algorithms.
With 24 cores, speedups over MKL are less dramatic, but Strassen (bottom left), $\bc{3}{3}{2}$ (bottom left), and $\bc{4}{3}{3}$ (bottom right) all outperform MKL and have similar performance.
Bini and Sch\"{o}hage have high performance, but they are APA algorithms and suffer from severe numerical problems.
}
\label{fig:parallel_square_perf}
\end{figure*}

\begin{figure*}[tb]
\centering
\includegraphics[width=3in]{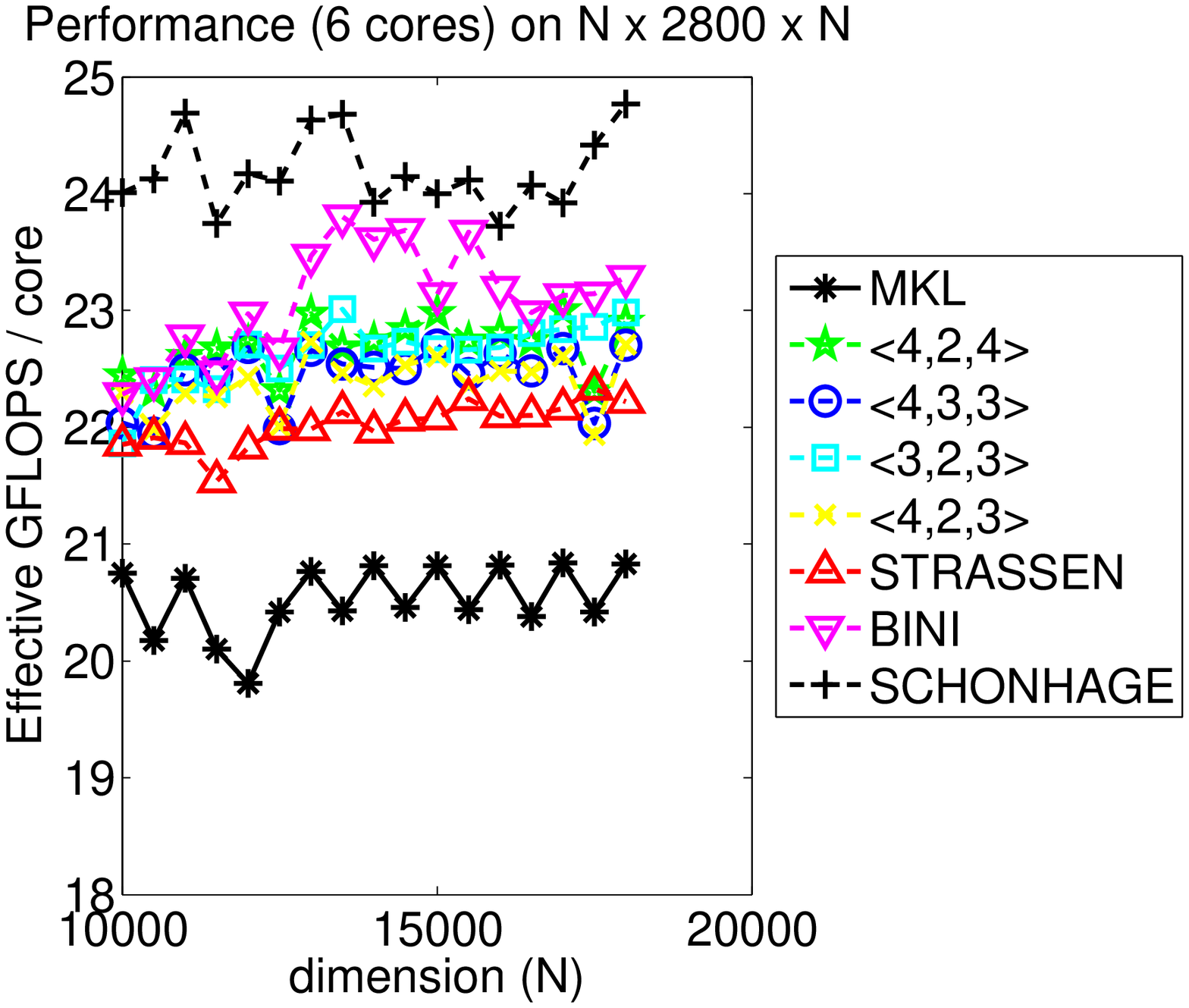}
\includegraphics[width=3in]{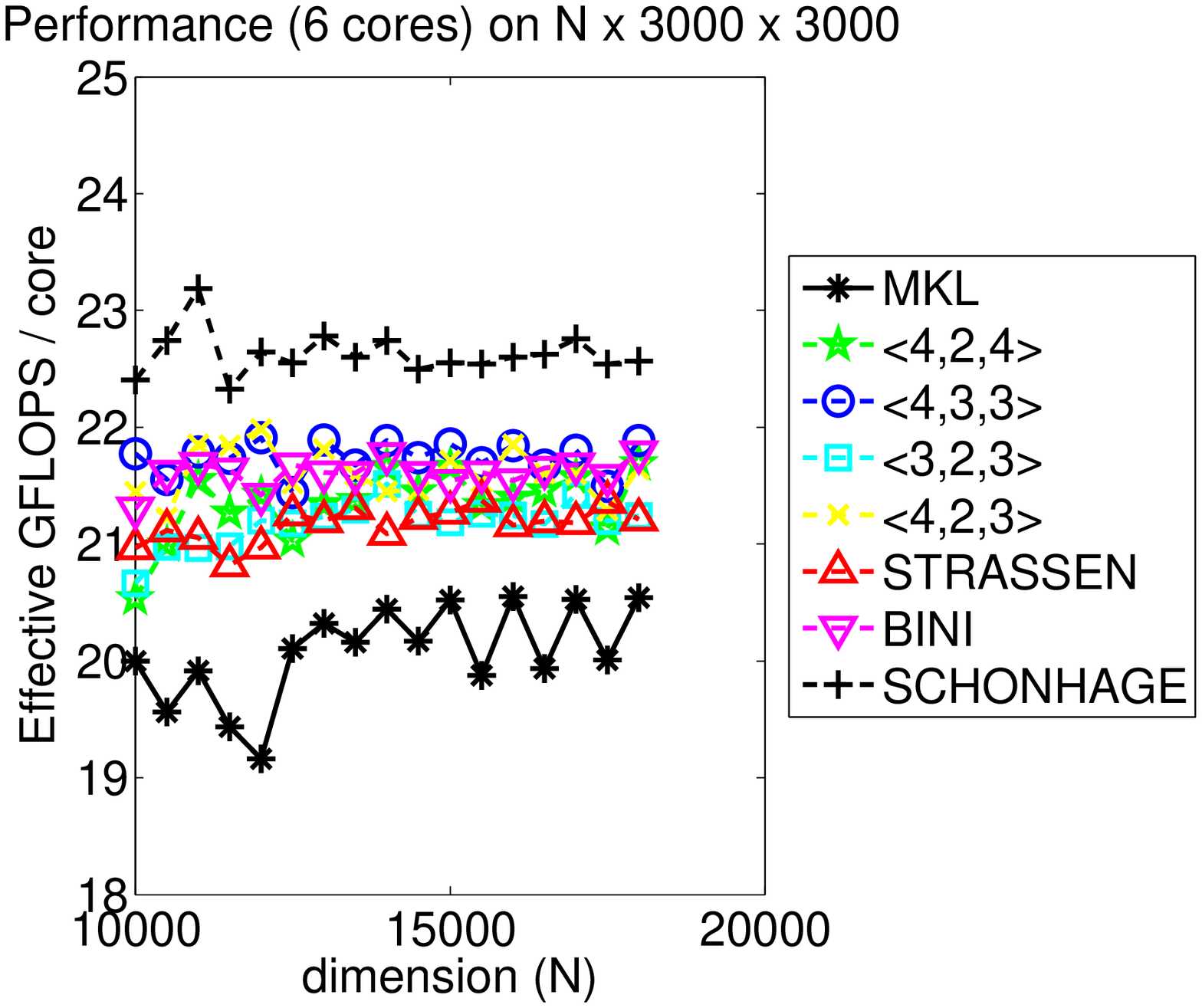} \\
\includegraphics[width=3in]{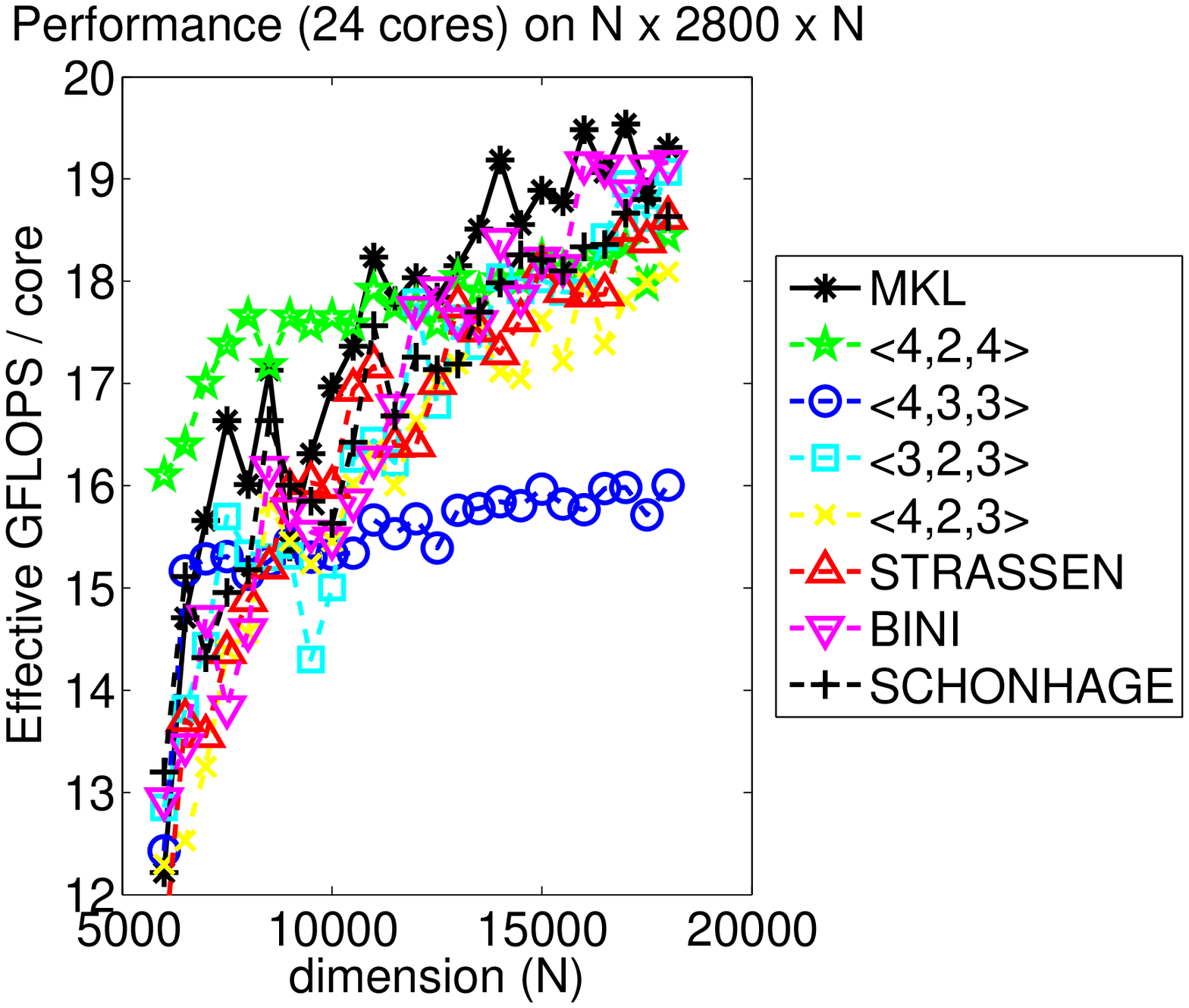}
\includegraphics[width=3in]{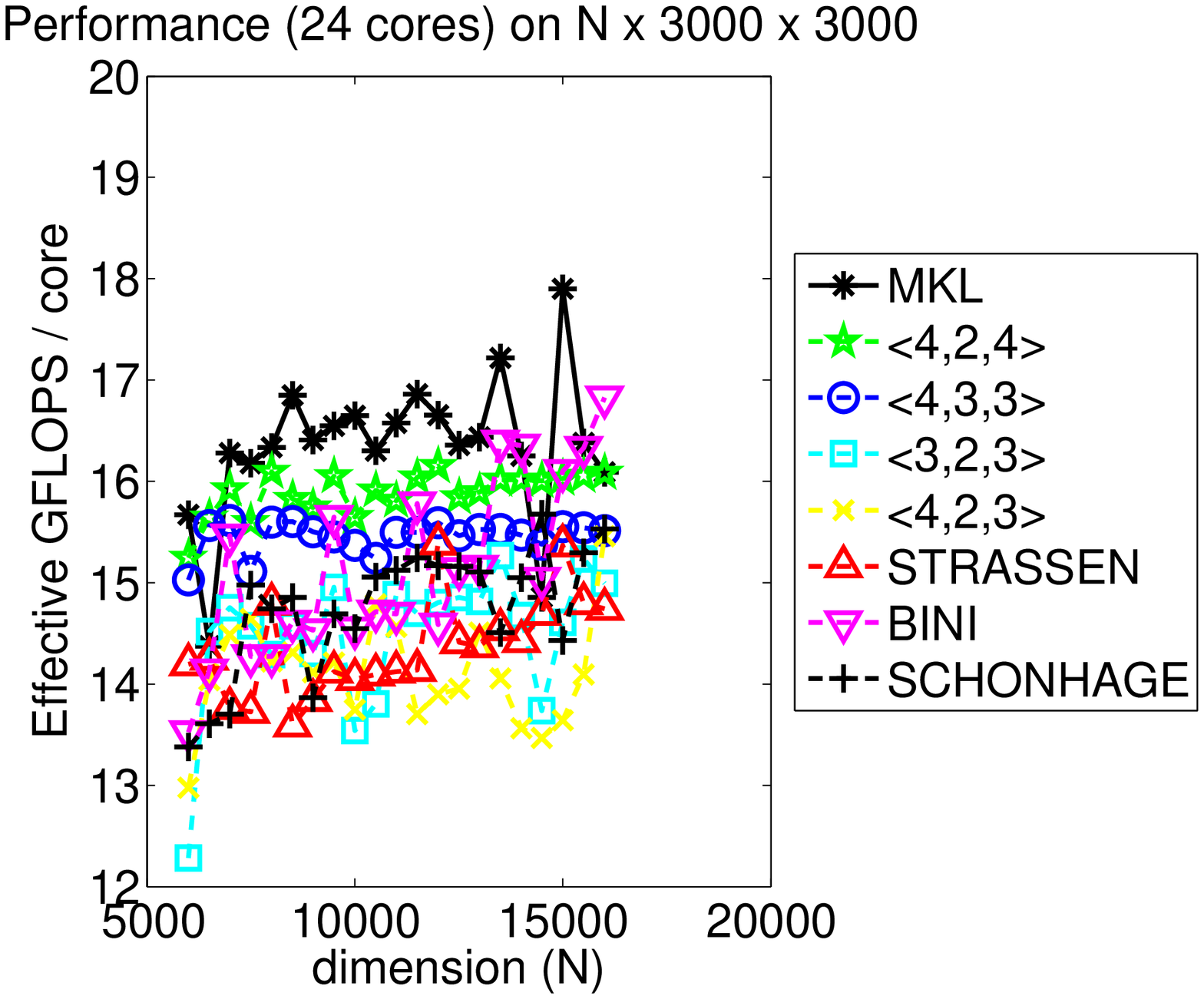}
\caption{
Effective parallel performance (Equation~\eqref{eqn:eff_perf}) of fast algorithms on rectangular problems using only 6 cores (top row) and all 24 cores (bottom row).
Problem sizes are an ``outer product" shape, $\dims{N}{2800}{N}$ (left column) and
multiplication of tall-and-skinny matrix by a small square matrix, $\dims{N}{3000}{3000}$ (right column).
With six cores, all fast algorithms outperform MKL, and new fast algorithms achieve about a 5\% performance gain over Strassen.
With 24 cores, bandwidth is a bottleneck and MKL outperforms fast algorithms.
}
\label{fig:parallel_nonsquare_perf}
\end{figure*}

Figure~\ref{fig:parallel_square_perf} shows the parallel performance for multiplying square matrices and Figure~\ref{fig:parallel_nonsquare_perf} shows the parallel performance for $\dims{N}{2800}{N}$ and $\dims{N}{3000}{3000}$ problem sizes.
We include performance on both 6 and 24 cores in order to illustrate the bandwidth issues discussed in Section~\ref{sec:bandwidth}.
We observe the following patterns in the parallel performance data:

\begin{enumerate}
\item
With 6 cores, bandwidth scaling is not a problem, and we find many of the same trends as in the sequential case.
All fast algorithms outperform MKL.
Apart from the APA algorithms, Strassen's algorithm is typically fastest for square matrices.
The $\bc{3}{2}{3}$ fast algorithm has the highest performance for the $\dims{N}{2800}{N}$ problem sizes, 
while $\bc{4}{3}{3}$ and $\bc{4}{2}{3}$ have the highest performance for the $\dims{N}{3000}{3000}$.
These algorithms match the shape of the problem.

\item
With 24 cores, MKL's $\dgemm$ is typically the highest performing algorithm for rectangular problem sizes (bottom row of Figure~\ref{fig:parallel_nonsquare_perf}).
In these problems, the ratio of time spent in matrix additions to time spent in matrix multiplication is too large, and bandwidth limitations prevent the fast algorithms from outperforming MKL.

\item 
With 24 cores and square problem sizes (bottom row of Figure~\ref{fig:parallel_square_perf}),
several algorithms outperform MKL.
Strassen's algorithm provides a modest speedup (around 5\%) and is one of highest performing exact algorithms.
The $\bc{4}{3}{3}$ and $\bc{4}{2}{4}$ fast algorithms outperform MKL and are competitive with Strassen.
The square problem sizes spend a large fraction of time in matrix multiplication, so the bandwidth costs for the matrix additions have less impact on performance.

\item
Again, the APA algorithms (Bini's algorithm and Sch\"{o}nhage's algorithm) have high performance on rectangular problem sizes.
It is still an open question if there exists a fast algorithm with the same complexity as Sch\"{o}nhage's algorithm.
Our results show that a significant performance gain is possible with such an algorithm.

\end{enumerate}

We also implemented the asymptotically fastest implementation of square matrix multiplication.
The algorithm is based on the $\bc{3}{3}{6}$ fast algorithm (Table~\ref{tab:algorithms}).
The square algorithm consists of composing $\bc{3}{3}{6}$, $\bc{3}{6}{3}$, $\bc{6}{3}{3}$ algorithms.
In other words, at the first recursive level, we use $\bc{3}{3}{6}$; at the second level $\bc{3}{6}{3}$; and at the third, $\bc{6}{3}{3}$.
The composed fast algorithm is for $\bc{3 \cdot 3 \cdot 6}{3 \cdot 6 \cdot 3}{6 \cdot 3 \cdot 3} = \bc{54}{54}{54}$.
Each step of the composed algorithm computes $40^3 = 64000$ matrix multiplications.
The asymptotic complexity of this algorithm is $\Theta(N^{\omega_0})$, with $\omega_0 = 3\log_{54}(40) \approx 2.775$.

Although this algorithm is \emph{asymptotically} the fastest, it does not perform well for the problem sizes considered in our experiments.
For example, with 6 cores and BFS parallelism, the algorithm achieved only 8.4 effective GFLOPS/core multiplying square with dimension $N = 13000$.
This is far below MKL's performance (Figure~\ref{fig:parallel_square_perf}).
We conclude that while the algorithm may be of theoretical interest, it does not perform well on the modest problem sizes of interest on shared memory machines.

\section{Discussion}

Our code generation framework lets us benchmark a large number of existing and new fast algorithms and test a variety of implementation details, such as how to handle matrix additions and how to implement the parallelism.
However, we performed only high-level optimizations; we believe more detailed tuning of fast algorithms can provide more performance gains.
Based on the performance results we obtain in this work, we can draw several conclusions in bridging the gap between the theory and practice of fast algorithms. 

First, in the case of multiplying square matrices, Strassen's algorithm consistently dominates the performance of exact algorithms (in sequential and parallel).
Even though the exact algorithm for $\bc{54}{54}{54}$ and Sch\"{o}nhage's APA algorithm\footnote{We note that the performance of Sch\"{o}nhage's APA algorithm accurately represents an exact algorithm for $\bc{3}{3}{3}$ with 21 multiplies, though it remains an open question whether such an exact algorithm exists.} for $\bc{3}{3}{3}$ are asymptotically faster in theory, they never outperform Strassen's for reasonable matrix dimensions in practice (sequential or parallel) because the overheads of the additions outweigh the reduction in multiplications.
This sheds some doubt on the prospect of finding a fast algorithm that will outperform Strassen's on square matrices; it will likely need to have a very small base case and still offer a significant reduction in multiplications.

On the other hand, another conclusion from our performance results is that for multiplying rectangular matrices (which typically occurs much more frequently than square in practice), there is a rich space for improvements.
In particular, fast algorithms with base cases that match the shape of the matrices tend to have the highest performance.
There are many promising algorithms in this space, and we suspect that algorithm-specific optimizations will prove fruitful.

Third, in the search for new fast algorithm, our results confirm an important metric.
Given a matrix multiplication tensor corresponding to base case $\bc{M}{K}{N}$, the rank of the decomposition $\alg{\M{U}}{\M{V}}{\M{W}}$ (\emph{i.e.}, the number of columns in each matrix) determines the exponent of the arithmetic complexity, and the number of nonzeros in the factor matrices determines the constant prefactor.
Our performance data demonstrates that for a given rank, minimizing the nonzeros in the factor matrices is indeed an important secondary goal.
Although the arithmetic cost associated with the sparsity of $\alg{\M{U}}{\M{V}}{\M{W}}$ is negligible in practice, the communication cost associated with each nonzero can be performance limiting.
We note that the communication costs of the streaming additions algorithm is independent of the sparsity, but the highest-performing additions algorithm in practice is the write-once algorithm, which is sensitive to the number of nonzeros.

Fourth, we have identified a parallel scaling impediment for fast algorithms, at least on shared memory architectures. 
Because the memory bandwidth often does not scale with the number of cores, and because the additions and multiplications are separate computations in our framework, the overhead of the additions compared to the multiplications worsens in the parallel case.
Short of fundamentally restructuring the fast algorithm implementations, this hardware bottleneck is unavoidable.
We note that on the distributed-memory architectures, this memory-bandwidth scaling bottleneck does not occur---the aggregate memory bandwidth scales with the number of nodes.

We would like to extend our framework to the distributed-memory case, in part because of the better prospects for parallel scaling.
A larger fraction of the time is spent in communication for the classical algorithm on this architecture, and fast algorithms can reduce the communication cost in addition to the computational cost in this case \cite{ballard2012communication}.
Similar code generation techniques will be helpful in exploring the performance of all the algorithms presented in this paper.

As matrix multiplication is the central computational kernel in linear algebra libraries, we would also like to incorporate these fast algorithms into frameworks like BLIS \cite{BLIS1} and PLASMA \cite{kurzak2013multithreading} to see how they affect a broader class of algorithms in numerical linear algebra.
We also plan to develop similar code generation techniques to explore fast algorithms on distributed-memory architectures.

Finally, we have not explored the numerical stability of the exact algorithms in order to compare their results.
While theoretical bounds can be derived from each algorithm's $\alg{\M{U}}{\M{V}}{\M{W}}$ representation, it is an open question which algorithmic properties are most influential in practice; our framework will allow for rapid empirical testing.
As numerical stability is an obstacle to widespread use of fast algorithms, extensive testing can help alleviate (or confirm) common concerns.


\acks

This research was supported in part by an appointment to the Sandia National Laboratories Truman Fellowship in National Security Science and Engineering, sponsored by Sandia Corporation (a wholly owned subsidiary of Lockheed Martin Corporation) as Operator of Sandia National Laboratories under its U.S. Department of Energy Contract No. DE-AC04-94AL85000.
Austin R. Benson is also supported by an Office of Technology Licensing Stanford Graduate Fellowship.

This research used resources of the National Energy Research Scientific Computing Center, which is supported by the Office of Science of the U.S. Department of Energy under Contract No. DE-AC02-05CH11231.


\bibliographystyle{abbrvnat}

\bibliography{bibliography}





\end{document}